\documentclass[12pt] {article}
\usepackage{psfrag}
\usepackage{graphicx}
\usepackage{latexsym,amsfonts}
\usepackage{amssymb}
\begin{document}
\setcounter{page}{1}
\def\theequation{\arabic{section}.\arabic{equation}}
\def\theequation{\thesection.\arabic{equation}}
\setcounter{section}{0}

\title{On kaonic hydrogen.\\ Quantum field theoretic and relativistic
 covariant approach}

\author{A. N. Ivanov\,\thanks{E--mail: ivanov@kph.tuwien.ac.at, Tel.:
+43--1--58801--14261, Fax: +43--1--58801--14299}~\thanks{Permanent
Address: State Polytechnical University, Department of Nuclear
Physics, 195251 St. Petersburg, Russian Federation}\,,
M. Cargnelli\,\thanks{E--mail: michael.cargnelli@oeaw.ac.at}\,,
M. Faber\,\thanks{E--mail: faber@kph.tuwien.ac.at, Tel.:
+43--1--58801--14261, Fax: +43--1--58801--14299},
J. Marton\,\thanks{E--mail: johann.marton@oeaw.ac.at}\,,\\
N. I. Troitskaya\,\thanks{State Polytechnical University, Department
of Nuclear Physics, 195251 St. Petersburg, Russian Federation},
J. Zmeskal\,\thanks{E--mail: johann.zmeskal@oeaw.ac.at}}

\date{\today}

\maketitle

\vspace{-0.5in}
\begin{center}
{\it Atominstitut der \"Osterreichischen Universit\"aten,
Arbeitsbereich Kernphysik und Nukleare Astrophysik, Technische
Universit\"at Wien, \\ Wiedner Hauptstr. 8-10, A-1040 Wien,
\"Osterreich \\ und\\ Institut f\"ur Mittelenergiephysik
\"Osterreichische Akademie der Wissenschaften,\\
Boltzmanngasse 3, A-1090, Wien}
\end{center}

\begin{center}
\begin{abstract}
We study kaonic hydrogen, the bound $K^- p$ state $A_{K p}$.  Within a
quantum field theoretic and relativistic covariant approach we derive
the energy level displacement of the ground state of kaonic hydrogen
in terms of the amplitude of $K^-p$ scattering for arbitrary relative
momenta.  The amplitude of low--energy $K^-p$ scattering near
threshold is defined by the contributions of three resonances
$\Lambda(1405)$, $\Lambda(1800)$ and $\Sigma^0(1750)$ and a smooth
elastic background. The amplitudes of inelastic channels of
low--energy $K^-p$ scattering fit experimental data on near threshold
behaviour of the cross sections and the experimental data by the DEAR
Collaboration. We use the soft--pion technique (leading order in
Chiral Perturbation Theory) for the calculation of the partial width
of the radiative decay of pionic hydrogen $A_{\pi p} \to n + \gamma$
and the Panofsky ratio. The theoretical prediction for the Panofsky
ratio agrees well with experimental data. We apply the soft--kaon
technique (leading order in Chiral Perturbation Theory) to the
calculation of the partial widths of radiative decays of kaonic
hydrogen $A_{Kp} \to \Lambda^0 + \gamma$ and $A_{K p} \to \Sigma^0 +
\gamma$. We show that the contribution of these decays to the width of
the energy level of the ground state of kaonic hydrogen is less than
1$\%$.
\end{abstract}


\end{center}

\newpage

\section{Introduction}
\setcounter{equation}{0}

Kaonic hydrogen $A_{K p}$ is an analogy of hydrogen with an electron
replaced by the $K^-$ meson. The relative stability of kaonic hydrogen
is fully due to Coulomb forces \cite{SD54}--\cite{IV2}.  The Bohr
radius of kaonic hydrogen is
\begin{eqnarray}\label{label1.1}
a_B = \frac{1}{\mu\,\alpha} = \frac{1}{\alpha}\,\Big(\frac{1}{m_{K^-}}
+ \frac{1}{m_p}\Big) = 83.594\,{\rm fm},
\end{eqnarray}
where $\mu = m_{K^-}m_p/( m_{K^-} + m_p) = 323.478\,{\rm MeV}$ is a
reduced mass of the $K^-p$ system, calculated at $m_{K^-} =
493.677\,{\rm MeV}$ and $m_p = 938.272\,{\rm MeV}$ \cite{DG00}, and
$\alpha = e^2/\hbar c = 1/137.036$ is the fine--structure constant
\cite{DG00}. Below we use the units $\hbar = c = 1$, then $\alpha =
e^2 = 1/137.036$. Since the Bohr radius of kaonic hydrogen is much
greater than the range of strong low--energy interactions $R_{\rm str}
\sim 1/m_{\pi^-} = 1.414\,{\rm fm}$, the strong low--energy
interactions can be taken into account perturbatively
\cite{SD54}--\cite{IV2}.

According to Deser, Goldberger, Baumann and Thirring \cite{SD54} the
energy level displacement of the ground state of kaonic hydrogen can
be defined in terms of the S--wave amplitude $f^{K^-p}_0(Q)$ of
low--energy $K^-p$ scattering as follows
\begin{eqnarray}\label{label1.2}
-\epsilon_{1s} + i\,\frac{\Gamma_{1s}}{2} =
 \frac{2\pi}{\mu}\,f^{K^-p}_0(0)\,|\Psi_{1s}(0)|^2,
\end{eqnarray}
where $\Psi_{1s}(0) = 1/\sqrt{\pi a^3_B}$ is the wave function of the
ground state of kaonic hydrogen at the origin and $f^{K^-p}_0(0)$ is
the amplitude of $K^-p$ scattering in the S--wave state, calculated at
zero relative momentum $Q = 0$ of the $K^-p$ pair. The DGBT formula
can be rewritten in the equivalent form
\begin{eqnarray}\label{label1.3}
-\epsilon_{1s} + i\,\frac{\Gamma_{1s}}{2} = 2\,\alpha^3 \mu^2\,
  f^{K^-p}_0(0),
\end{eqnarray}
where $2\,\alpha^3 \mu^2 = 412.124\,{\rm eV\,fm^{-1}}$ and
$f^{K^-p}_0(0)$ is measured in ${\rm fm}$. The formula
(\ref{label1.3}) is used by experimentalists for the analysis of
experimental data on the energy level displacement of the ground state
of kaonic hydrogen \cite{DEAR1}--\cite{DEAR4}.

For non--zero relative momentum $Q$ the amplitude $f^{K^-p}_0(Q)$ is
defined by
\begin{eqnarray}\label{label1.4}
f^{K^-p}_0(Q) = \frac{1}{2iQ}\,\Big(\eta^{K^-p}_0(Q)\,
e^{\textstyle\,2i\delta^{K^-p}_0(Q)} - 1\Big),
\end{eqnarray}
where $\eta^{K^-p}_0(Q)$ and $\delta^{K^-p}_0(Q)$ are the inelasticity
and the phase shift of the reaction $K^- + p \to K^- + p$,
respectively. At relative momentum zero, $Q = 0$, the inelasticity and
the phase shift are equal to $\eta^{K^-p}_0(0) = 1$ and
$\delta^{K^-p}_0(0) = 0$. For $Q \to 0$ the phase shift behaves as
$\delta^{K^-p}_0(Q) = a^{K^-p}_0\,Q + O(Q^2)$, where $a^{K^-p}_0$ is
the S--wave scattering length of $K^-p$ scattering.

The real part of $f^{K^-p}_0(0)$ is related to $a^{K^-p}_0$ as
\begin{eqnarray}\label{label1.5}
{\cal R}e\,f^{K^-p}_0(0) = a^{K^-p}_0 = \frac{1}{2}\,(a^0_0 +
a^1_0),
\end{eqnarray}
where $a^0_0$ and $a^1_0$ are the S--wave scattering lengths $a^I_0$
with isospin $I = 0$ and $I = 1$, respectively.

Due to the optical theorem the imaginary part of the amplitude
$f^{K^-p}_0 (0)$ is related to the total cross section
$\sigma^{K^-p}_0(Q)$ for $K^-p$ scattering in the S--wave state
\begin{eqnarray}\label{label1.6}
{\cal I}m\,f^{K^-p }_0(0) = \lim_{Q \to
0}\frac{Q}{4\pi}\,\sigma^{K^-p}_0(Q) = \frac{1}{2}\lim_{Q \to
0}\frac{1}{Q}\,(1 - \eta^{K^-p}_0(Q)\cos 2\delta^{K^-p}_0(Q)).
\end{eqnarray}
 The r.h.s. of (\ref{label1.6}) can be transcribed into the form
\begin{eqnarray}\label{label1.7}
{\cal I}m\,f^{K^-p}_0(0) =
-\frac{1}{2}\,\frac{d\eta^{K^-p}_0(Q)}{dQ}\Big|_{Q = 0}.
\end{eqnarray}
Hence, according to the DGBT formula the energy level displacement of
the ground state of kaonic hydrogen is defined by
\begin{eqnarray}\label{label1.8}
\epsilon_{1s} &=&- 2\,\alpha^3 \mu^2\, {\cal R}e\,f^{K^-p}_0(0) = -
  2\,\alpha^3 \mu^2\, a^{K^-p}_0,\nonumber\\ \Gamma_{1s} &=& ~~4
  \,\alpha^3 \mu^2\,{\cal I}m\,f^{K^-p}_0(0) = - 2 \,\alpha^3 \mu^2\,
  \frac{d\eta^{K^-p}_0(Q)}{dQ}\Big|_{Q = 0}.
\end{eqnarray}
The recent preliminary experimental data on the energy level
displacement of the ground state of kaonic hydrogen obtained by the
DEAR Collaboration \cite{DEAR4} read
\begin{eqnarray}\label{label1.9}
- \epsilon^{\exp}_{1s} + i\,\frac{~\Gamma^{\exp}_{1s}}{2} = (-183\pm
  62) + i\,(106 \pm 69)\;{\rm eV}.
\end{eqnarray}
In this paper we give (i) a model--independent, quantum field
theoretic and relativistic covariant derivation of the energy level
displacement of the ground state of kaonic hydrogen and (ii) a
theoretical modeling of the amplitude of $K^-p$ scattering in the
S--wave state $f^{K^-p}_0(Q)$ near threshold of the $K^-p$ pair $Q
\approx 0$, fitting well experimental data (\ref{label1.9}) by the
DEAR Collaboration \cite{DEAR4}.

The paper is organized as follows. In Section 2 we write down the wave
function of the ground state of kaonic hydrogen within the quantum
field theoretic and relativistic covariant approach developed in
\cite{IV1,IV2} (see also \cite{CD03}). In Section 3 we derive the
energy level displacement of the ground state of kaonic hydrogen in a
model--independent way.  In Section 4 we describe the amplitude of
$K^-p$ scattering near threshold by the contributions of the
resonances $\Lambda(1405)$, $\Lambda(1800)$ and $\Sigma(1750)$. The
obtained amplitude of $K^-p$ scattering we use for the calculation of
the energy level displacement of the ground state of kaonic
hydrogen. In Section 5 we calculate the contribution of the elastic
background to the amplitude of low--energy $K^-p$ scattering. We show
that the theoretical results fit well preliminary experimental data by
the DEAR Collaboration \cite{DEAR4}. In Section 6 we calculate the
partial widths of the radiative decay channels of kaonic hydrogen
$A_{Kp} \to \Lambda^0 + \gamma$ and $A_{Kp} \to \Sigma^0 + \gamma$
\cite{JG03}. First, we develop technique and methodics, based on the
soft--pion(kaon) technique, for the calculation of the partial width
of the decay $A_{\pi p} \to n + \gamma$ of pionic hydrogen in the
ground state. We calculate the Panofsky ratio, $1/P = \Gamma(A_{\pi p}
\to n + \gamma)/\Gamma((A_{\pi p} \to n + \pi^0) = 0.681 \pm 0.048$,
in agreement with the experimental value $1/P = 0.647 \pm 0.004$
\cite{JS77}. The application of this technique to the calculation of
the partial widths of the decays $A_{Kp} \to \Lambda^0 + \gamma$ and
$A_{Kp} \to \Sigma^0 + \gamma$ shows that the contribution of these
decay channels to the width of the energy level of the ground state of
kaonic hydrogen is less than 1$\%$. In the Conclusion we discuss the
obtained results. We show that our approach to the description of
low--energy $K^-p$ scattering is consistent with the experimental data
by the DEAR Collaboration \cite{DEAR4}.  In the Appendix we calculate
the elastic background of S--wave elastic $K^-p$ scattering near
threshold within the Effective quark model with chiral $U(3)\times
U(3)$ symmetry \cite{AI99}--\cite{AI92}.

\section{Ground state wave function of kaonic hydrogen}
\setcounter{equation}{0}

The wave function of kaonic hydrogen in the ground state we define as
\cite{IV1,IV2,SS61,IZ80}
\begin{eqnarray}\label{label2.1}
|A^{(1s)}_{K p}(\vec{P},\sigma_p)\rangle &=& \frac{1}{(2\pi)^3}\int
 \frac{d^3k_{K^-}}{\sqrt{2E_{K^-}(\vec{k}_{K^-})}}
 \frac{d^3k_p}{\sqrt{2E_p(\vec{k}_p)}} \delta^{(3)}(\vec{P} -
 \vec{k}_{K^-} - \vec{k}_p)\,\nonumber\\
 &&\times\,\sqrt{2E^{(1s)}_A(\vec{k}_{K^-} +\vec{k}_p) }\,
 \Phi_{1s}(\vec{k}_{K^-})|K^-(\vec{k}_{K^-})
 p(\vec{k}_p,\sigma_p)\rangle,
\end{eqnarray}
where $E^{(1s)}_A(\vec{P}\,) = \sqrt{{M^{(1s)}_A}^{\textstyle _2} +
\vec{P}^{\;2}}$ and $\vec{P}$ are total energy and momentum of kaonic
hydrogen, $M^{(1s)}_A = m_p + m_{K^-} + E_{1s}$ and $E_{1s} = -
8613\,{\rm eV}$ are mass and binding energy of kaonic hydrogen in the
ground bound state, $\sigma_p$ is a polarization of the proton. Then,
$\Phi_{1s}(\vec{k}_{K^-})$ is the wave function of the ground state in
the momentum representation normalized by
\begin{eqnarray}\label{label2.2}
\int \frac{d^3k}{(2\pi)^3}\,|\Phi_{1s}(\vec{k}\,)|^2 = 1.
\end{eqnarray}
The wave function $|K^-(\vec{k}_{K^-}) p(\vec{k}_p,\sigma_p)\rangle$
 we define as \cite{IV1,IV2,SS61,IZ80}
\begin{eqnarray}\label{label2.3}
|K^-(\vec{k}_{K^-})p(\vec{k}_p,\sigma_p)\rangle =
 c^{\dagger}_{K^-}(\vec{k}_{K^-})a^{\dagger}_p(\vec{k}_p,
 \sigma_p)|0\rangle,
\end{eqnarray}
where $c^{\dagger}_{K^-}(\vec{k}_{K^-})$ and $a^{\dagger}_p(\vec{k}_p,
\sigma_p)$ are operators of creation of the $K^-$ meson with momentum
$\vec{k}_{K^-}$ and the proton with momentum $\vec{k}_p$ and
polarization $\sigma_p = \pm 1/2$. They satisfy standard relativistic
covariant commutation and anti--commutation relations \cite{IV1,SS61}.
The wave function (\ref{label2.1}) is normalized by
\begin{eqnarray}\label{label2.4}
\langle A^{(1s)}_{K p}(\vec{P}\,',\sigma\,'_p)|A^{(1s)}_{K
p}(\vec{P},\sigma_p)\rangle &=& (2\pi)^3\,
2E^{(1s)}_A(\vec{P}\,)\,\delta^{(3)}(\vec{P}\,' -
\vec{P}\,)\,\delta_{\sigma\,'_p\sigma_p}\int
\frac{d^3k}{(2\pi)^3}\,|\Phi_{1s}(\vec{k}\,|^2 = \nonumber\\ &=&
(2\pi)^3\, 2E^{(1s)}_A(\vec{P}\,)\,\delta^{(3)}(\vec{P}\,' -
\vec{P}\,)\,\delta_{\sigma\,'_p\sigma_p}.
\end{eqnarray}
This is a relativistic covariant normalization of the wave
function. 

The wave function (\ref{label2.1}) we will apply to the calculation of
the energy level displacement of the ground state of kaonic hydrogen
within a quantum field theoretic and relativistic covariant approach.

\section{Energy level displacement of the ground state}
\setcounter{equation}{0}

According to \cite{IV1,IV2,SS61}, the energy level displacement of the
ground state of kaonic hydrogen is defined by
\begin{eqnarray}\label{label3.1}
-\,\epsilon_{1s} + i\,\frac{\Gamma_{1s}}{2} = \lim_{T,V\to
\infty}\frac{\langle A^{(1s)}_{K
p}(\vec{P},\sigma_p)|\mathbb{T}|A^{(1s)}_{K
p}(\vec{P},\sigma_p)\rangle}{2 E^{(1s)}_A(\vec{P}\,)VT}\Big|_{\vec{P}
= 0},
\end{eqnarray}
where $TV$ is a 4--dimensional volume defined by
$(2\pi)^4\delta^{(4)}(0) = TV$ \cite{SS61} and $\mathbb{T}$ is the
$T$--matrix obeying the unitary condition \cite{SS61,IZ80}
\begin{eqnarray}\label{label3.2}
\mathbb{T} - \mathbb{T}^{\dagger} = i\,\mathbb{T}^{\dagger}\mathbb{T}.
\end{eqnarray}
Using the wave function (\ref{label2.1}) we reduce the r.h.s. of
(\ref{label3.1}) to the form
\begin{eqnarray}\label{label3.3}
\hspace{-0.3in}-\,\epsilon_{1s} + i\,\frac{\Gamma_{1s}}{2} &=&
\frac{1}{4m_{K^-}m_p}\int\frac{d^3k}{(2\pi)^3}
\int\frac{d^3q}{(2\pi)^3}\,
\sqrt{\frac{m_{K^-}m_p}{E_{K^-}(\vec{k}\,)E_p(\vec{k}\,)}}\,
\sqrt{\frac{m_{K^-}m_p}{E_{K^-}(\vec{q}\,) E_p(\vec{q}\,)}}\nonumber\\
\hspace{-0.3in}&&\times\,\Phi^{\dagger}_{1s}(\vec{k}\,)\,\lim_{T,V\to
\infty}\frac{\langle
K^-(\vec{k}\,)p(-\vec{k},\sigma_p)|\mathbb{T}|K^-(\vec{q}\,)p(-\vec{q},\sigma_p)\rangle}{VT}\,
\Phi_{1s}(\vec{q}\,),
\end{eqnarray}
where the matrix element of the $T$--matrix defines the amplitude of
$K^-p$ scattering\,\footnote{In Chiral Perturbation Theory (ChPT)
\cite{JG99,JG83} the $T$--matrix can be expressed in terms of an
effective Lagrangian ${\cal L}_{\rm eff}(x)$ (see also
\cite{IV1,IV2}). If all loop--contributions are taken into account and
renormalization is carried out the effective Lagrangian ${\cal L}_{\rm
eff}(x)$ can be used only in the tree--approximation \cite{RD69} (see
also \cite{IV1,IV2}).}
\begin{eqnarray}\label{label3.4}
&&\lim_{T,V\to \infty}\frac{\langle
K^-(\vec{k}\,)p(-\vec{k},\sigma_p)|\mathbb{T}
|K^-(\vec{q}\,)p(-\vec{q},\sigma_p)\rangle}{VT} = \nonumber\\
&&\hspace{1in} = M(K^-(\vec{q}\,)p(-\vec{q},\sigma_p) \to
K^-(\vec{k}\,)p(-\vec{k},\sigma_p)).
\end{eqnarray}
Thus, the energy level displacement of the ground state of kaonic
hydrogen is defined by the amplitude of $K^-p$ scattering
\cite{IV1,IV2}
\begin{eqnarray}\label{label3.5}
- \epsilon_{1s} + i\,\frac{\Gamma_{1s}}{2} &=&
  \frac{1}{4m_{K^-}m_p}\int\frac{d^3k}{(2\pi)^3}
  \int\frac{d^3q}{(2\pi)^3}\,
  \sqrt{\frac{m_{K^-}m_p}{E_{K^-}(\vec{k}\,)E_p(\vec{k}\,)}}\,
  \sqrt{\frac{m_{K^-}m_p}{E_{K^-}(\vec{q}\,)
  E_p(\vec{q}\,)}}\nonumber\\ &\times&\Phi^{\dagger}_{1s}(\vec{k}\,)\,
  M(K^-(\vec{q}\,)p(-\vec{q},\sigma_p) \to
  K^-(\vec{k}\,)p(-\vec{k},\sigma_p))\, \Phi_{1s}(\vec{q}\,),
\end{eqnarray}
Due to the wave functions $\Phi^{\dagger}_{1s}(\vec{k}\,)$ and
$\Phi_{1s}(\vec{q}\,)$ the main contributions to the integrals over
$\vec{k}$ and $\vec{q}$ come from the regions of 3--momenta $k \sim
1/a_B$ and $q \sim 1/a_B$, where $1/a_B = 2.361\,{\rm MeV}$. Since
typical momenta in the integrand are much less than the masses of
coupled particles, $m_{K^-} \gg 1/a_B$ and $m_p \gg 1/a_B$, the
amplitude of $K^-p$ scattering can be defined for low--energy momenta
only\,\footnote{It is obvious that due to the formula (\ref{label3.5})
a knowledge of the amplitude of $K^-p$ scattering for all relative
momenta from zero to infinity should give a possibility to calculate
the energy level displacement of the ground state of kaonic hydrogen
without any low--energy approximation.}.

Following \cite{IV1,IV2} the amplitude of low--energy $K^-p$
scattering we define as 
\begin{eqnarray}\label{label3.6}
M(K^-(\vec{q}\,)p(-\vec{q},\sigma_p) \to
  K^-(\vec{k}\,)p(-\vec{k},\sigma_p)) = 8\pi\,(m_{K^-} +
  m_p)\,f^{K^-p}_0(\sqrt{kq}),
\end{eqnarray}
where the amplitude $f^{K^-p}_0(\sqrt{kq})$ is determined by
\begin{eqnarray}\label{label3.7}
f^{K^-p}_0(\sqrt{kq}) =
\frac{1}{2i\sqrt{kq}}\,\Big(\eta^{K^-p}_0(\sqrt{kq})\,
e^{\textstyle\,2i\delta^{K^-p}_0(\sqrt{kq})} - 1\Big).
\end{eqnarray}
The shift and width of the energy level of the ground state of kaonic
hydrogen are equal to
\begin{eqnarray}\label{label3.8}
\epsilon_{1s} &=& - \frac{2\pi}{\mu}
\int\!\!\!\int\frac{d^3k}{(2\pi)^3}\, \frac{d^3q}{(2\pi)^3}\,
\sqrt{\frac{m_{K^-}m_p}{E_{K^-}(\vec{k}\,)\,
E_p(\vec{k}\,)}}\,\sqrt{\frac{m_{K^-}m_p}{E_{K^-}(\vec{q}\,)
E_p(\vec{q}\,)}}\Phi^{\dagger}_{1s}(\vec{k}\,)\,\Phi_{1s}(\vec{q}\,)
\nonumber\\ &&\times\, \eta^{K^-p}_0(\sqrt{kq})\,\frac{\sin
2\delta^{K^-p}_0(\sqrt{kq})}{2\sqrt{kq}},\nonumber\\
\Gamma_{1s}&=&\frac{2\pi}{\mu} \int\!\!\!\int\frac{d^3k}{(2\pi)^3}\,
\frac{d^3q}{(2\pi)^3}\, \sqrt{\frac{m_{K^-}m_p}{E_{K^-}(\vec{k}\,)
E_p(\vec{k}\,)}}\,\sqrt{\frac{m_{K^-}m_p}{E_{K^-}(\vec{q}\,)
E_p(\vec{q}\,)}}\,\Phi^{\dagger}_{1s}(\vec{k}\,)\,
\Phi_{1s}(\vec{q}\,)\, \nonumber\\ &&\times\,\frac{1}{\sqrt{kq}}\,(1 -
\eta^{K^-p}_0(\sqrt{kq})\,\cos 2\delta^{K^-p}_0(\sqrt{kq}))
=\nonumber\\ &=&\frac{1}{2\mu} \int\!\!\!\int\frac{d^3k}{(2\pi)^3}\,
\frac{d^3q}{(2\pi)^3}\, \sqrt{\frac{m_{K^-}m_p}{E_{K^-}(\vec{k}\,)
E_p(\vec{k}\,)}}\,\sqrt{\frac{m_{K^-}m_p}{E_{K^-}(\vec{q}\,)
E_p(\vec{q}\,)}}\,\Phi^{\dagger}_{1s}(\vec{k}\,)\,
\Phi_{1s}(\vec{q}\,)\, \nonumber\\
&&\times\,\sqrt{kq}\,\sigma^{K^-p}_0(\sqrt{kq}).
\end{eqnarray}
The formula (\ref{label3.8}) reduces to the the DGBT formula defining
the amplitude of $K^-p$ scattering at $k = q = 0$ \cite{IV1,IV2}.  We
would like to emphasize that the main contributions to the momentum
integrals in (\ref{label3.8}) comes from the region $k \sim q \sim
1/a_B = 2.361\,{\rm MeV}$ but not from $k = q = 0$. Hence, the
calculation of the amplitude of $K^-p$ scattering at $k = q = 0$ is
not an explicit result but an approximation, which is well--defined
only if the amplitude of $K^-p$ scattering is a smooth function near
threshold\,\footnote{Practically, the corrections to the energy level
displacement, coming from a momentum expansion of the amplitude of
$K^-p$ scattering, are of order of powers of $\alpha$. This means that
the term of order $O(Q)$ gives a correction of order of $O(\alpha)$,
multiplied by the derivative of the amplitude of $K^-p$ scattering
with respect to the relative momentum $Q$, calculated at $Q = 0$. The
convergence of this expansion is fully defined by the derivatives of
the amplitude of $K^-p$ scattering. Such corrections, caused by
Coulombic photons, should be taken into account on the same footing as
the corrections caused by QCD isospin--breaking and electromagnetic
interactions \cite{JG02,TE03} (see also \cite{TE88}).}.

Assuming that near threshold the amplitude of low--energy $K^-p$
scattering is a smooth function of relative momentum $Q$ of the $K^-p$
pair and keeping only the leading terms in momentum expansion at $Q =
0$, we arrive at the energy level displacement of the ground state of
kaonic hydrogen
\begin{eqnarray}\label{label3.9}
- \epsilon_{1s} + i\,\frac{\Gamma_{1s}}{2}=
\frac{2\pi}{\mu}\,\Big[a^{K^-p}_0 -
i\,\frac{1}{2}\,\frac{d\eta^{K^-p}_0(Q)}{dQ}\Big|_{Q = 0}\Big]\Bigg|
\int\frac{d^3k}{(2\pi)^3}\,
\sqrt{\frac{m_{K^-}m_p}{E_{K^-}(\vec{k}\,)\,
E_p(\vec{k}\,)}}\,\Phi_{1s}(\vec{k}\,)\Bigg|^2.
\end{eqnarray}
This is the quantum field theoretic, relativistic covariant and
model--independent generalization of the DGBT formula (\ref{label1.2})
\cite{IV1,IV2}.

The amplitude of low--energy $K^-p$ scattering we represent in the
form 
\begin{eqnarray}\label{label3.10}
f^{K^-p}_0(Q) &=& \frac{1}{2iQ}\,\Big(\eta^{K^-p}_0(Q)\,
e^{\textstyle\,2i\delta^{K^-p}_0(Q)} - 1\Big) =\nonumber\\ &=&
\frac{1}{2iQ}\,\Big( e^{\textstyle\,2i\delta^{K^-p}_B(Q)} - 1\Big) +
e^{\textstyle\,2i\delta^{K^-p}_B(Q)}f^{K^-p}_0(Q)_R,
\end{eqnarray}
where $\delta^{K^-p}_0(Q)_B$ is the phase shift of an elastic
background of low--energy $K^-p$ scattering and $f^{K^-p}_0(Q)_R$ is
the contribution of resonances.

We assume that $f^{K^-p}_0(Q)_R$ is defined by the contributions of
the $\Lambda(1405)$ resonance, an $SU(3)_{\rm flavour}$ singlet
\cite{DG00a}, and the $\Lambda(1800)$ and $\Sigma(1750)$ resonances,
components of the $SU(3)_{\rm flavour}$ octet
\cite{DG00b}\,\footnote{Recall, that the resonances $\Lambda(1405)$
and $\Lambda(1800)$ have the status $****$, whereas the $\Sigma(1750)$
resonance has a status $***$ \cite{DG00a,DG00b}.}. For simplicity we
denote $\Lambda(1405)$ as $\Lambda^0_1$ and $\Lambda(1800)$ and
$\Sigma(1750)$ as $\Lambda^0_2$ and $\Sigma^0_2$\,\footnote{We keep
only the neutral component of the $\Sigma(1750)$ resonance.}.

\section{Amplitude of low--energy $K^-p$ scattering. Resonances}
\setcounter{equation}{0}

Treating the resonances $\Lambda(1405)$, $\Lambda(1800)$ and
$\Sigma(1750)$ as elementary fields\,\footnote{This agrees, for
instance, with the approach developed within ChPT in \cite{MR96}.}
we can write down phenomenological interactions
\begin{eqnarray}\label{label4.1}
{\cal L}_{\Lambda_1BP}(x) &=& g_1\bar{\Lambda}^0_1(x)\,{\rm
tr}\{B(x)P(x)\} + {\rm h.c.} = g_1\bar{\Lambda}_1(x)\,B^b_a(x)P^a_b(x)
+ {\rm h.c.}, \nonumber\\ {\cal L}_{B_2BP}(x) &=&
\frac{1}{\sqrt{2}}\,g_2\,{\rm tr}\{\{\bar{B}_2,B\}P\} +
\frac{1}{\sqrt{2}}\,f_2\,{\rm tr}\{[\bar{B}_2,B]P\} + {\rm h.c.} =
\nonumber\\ &=&\frac{1}{\sqrt{2}}\,(g_2 + f_2)\,(\bar{B}_2)^b_a
B^a_cP^c_b + \frac{1}{\sqrt{2}}\,(g_2 - f_2)\,(\bar{B}_2)^b_a B^c_b
P^a_c + {\rm h.c.},
\end{eqnarray}
where $g_1$, $g_2$ and $f_2$ are phenomenological coupling constants,
$\Lambda^0_1(x)$, $(\bar{B}_2)^b_a(x)$, $B^b_a(x)$ and
$P^a_b(x)$\,($a(b) = 1,2,\ldots,8$) are interpolating fields of the
$\Lambda(1405)$--resonance, the octet of baryon resonances
$\Lambda(1800)$ and $\Sigma(1750)$, the octet of light baryons and the
octet of pseudoscalar mesons, respectively:
\begin{eqnarray}\label{label4.2}
(\bar{B}_2)^b_a &=& \left(\begin{array}{llcl} {\displaystyle
\frac{\bar{\Sigma}^0_2}{\sqrt{2}} +
\frac{\bar{\Lambda}^0_2}{\sqrt{6}}} &
\hspace{0.3in}\bar{\Sigma}^-_2 & - \bar{\Xi}^-_2 \\
\hspace{0.3in}\bar{\Sigma}^+_2 &{\displaystyle -
\frac{\bar{\Sigma}^0_2}{\sqrt{2}} +
\frac{\bar{\Lambda}^0_2}{\sqrt{6}}} & \bar{\Xi}^0_2 \\
\hspace{0.3in}\bar{p}_2& \hspace{0.3in} \bar{n}_2 & {\displaystyle
-\frac{2}{\sqrt{6}}\bar{\Lambda}^0_2} \\
\end{array}\right),\nonumber\\
B^b_a &=& \left(\begin{array}{llcl} {\displaystyle
\frac{\Sigma^0}{\sqrt{2}} + \frac{\Lambda^0}{\sqrt{6}}} &
\hspace{0.3in}\Sigma^+ & p \\ \hspace{0.3in}\Sigma^- &{\displaystyle -
\frac{\Sigma^0}{\sqrt{2}} + \frac{\Lambda^0}{\sqrt{6}}} & n \\
\hspace{0.15in}- \Xi^- & \hspace{0.3in}\Xi^0 & {\displaystyle
-\frac{2}{\sqrt{6}}\Lambda^0}
\end{array}\right),\nonumber\\
P^a_b &=& \left(\begin{array}{llcl} {\displaystyle
\frac{\pi^0}{\sqrt{2}} + \frac{\eta}{\sqrt{6}}} &
\hspace{0.3in} \pi^+ & K^+\\ \hspace{0.3in} \pi^-
&{\displaystyle - \frac{\pi^0}{\sqrt{2}} +
\frac{\eta}{\sqrt{6}}} & K^0\\
\hspace{0.15in}- K^- & \hspace{0.3in} \bar{K}^0& {\displaystyle
-\frac{2}{\sqrt{6}}\,\eta} \\
\end{array}\right).
\end{eqnarray}
For simplicity we identify the component $\eta(x)$ of the pseudoscalar
octet with the observed pseudoscalar meson $\eta(550)$ \cite{DG00}.

Keeping only terms relevant to low--energy $K^-p$ scattering we reduce
the effective Lagrangians (\ref{label4.1}) to the form
\begin{eqnarray}\label{label4.3}
{\cal L}_{\Lambda^0_1BP}(x) &=&
g_1\,\bar{\Lambda}^0_1(x)(\vec{\Sigma}(x)\cdot \vec{\pi}(x) -
p(x)K^-(x) + n(x)\bar{K}^0(x) + \frac{1}{3}\,\Lambda^0(x)\eta(x)) +
{\rm h.c.}\nonumber\\ {\cal L}_{\Lambda^0_2BP}(x) &=&
\frac{g_2}{\sqrt{3}}\,\bar{\Lambda}^0_2(x)(\vec{\Sigma}(x)\cdot
\vec{\pi}(x) - \Lambda^0(x)\eta(x))\nonumber\\ &+& \frac{g_2 +
3f_2}{2\sqrt{3}}\,\bar{\Lambda}^0_2(x)\,(p(x)K^-(x) -
n(x)\bar{K}^0(x)) + {\rm h.c.},\nonumber\\ {\cal
L}_{\Sigma^0_2BP}(x)&=& f_2\,\bar{\Sigma}^0_2(x)\,(\Sigma^-(x)\pi^+(x)
- \Sigma^+(x)\pi^-(x))\nonumber\\ &+&
\frac{g_2}{\sqrt{3}}\,\bar{\Sigma}^0_2(x)\,(\Lambda^0(x)\pi^0(x) +
\Sigma^0(x)\eta(x))\nonumber\\ &+& \frac{g_2 -
f_2}{2}\,\bar{\Sigma}^0_2(x)\,(- p(x)K^-(x) - n(x)\bar{K}^0(x)) + {\rm
h.c.}.
\end{eqnarray}
According to (\ref{label3.10}) at threshold $Q = 0$ the amplitude
$f^{K^-p}_0(0)$ of $K^-p$ scattering we define as
\begin{eqnarray}\label{label4.4}
f^{K^-p}_0(0) = A^{K^-p}_B + f^{K^-p}_0(0)_R,
\end{eqnarray}
where $A^{K^-p}_B$ is a real parameter\,\footnote{We calculate the
parameter $A^{K^-p}_B$ in Section 5.}, describing a smooth elastic
background $\delta^{K^-p}_0(Q)_B = A^{K^-p}_BQ$, and $f^{K^-p}_0(0)_R$
is the contribution of the resonances, which we determine as

\begin{eqnarray}\label{label4.5}
f^{K^-p}_0(0)_R = \frac{1}{2}\,\Big(f^{K^-p}_0(0)_{I = 0} +
f^{K^-p}_0(0)_{I = 1}\Big),
\end{eqnarray}
where the amplitudes $f^{K^-p}_0(0)_{I = 0}$ and $f^{K^-p}_0(0)_{I =
1}$ of low--energy $K^-p$ scattering with isospin $I = 0$ and isospin
$I = 1$ are saturated by the $\Lambda(1405)$, $\Lambda(1800)$ and
$\Sigma(1750)$ resonances, respectively.  The amplitude
$f^{K^-p}_0(0)_R$ contains real and imaginary parts ${\cal
R}e\,f^{K^-p}_0(0)_R$ and ${\cal I}m\,f^{K^-p}_0(0)_R$, which define
elastic and inelastic channels.

\subsection{Imaginary part of $f^{K^-p}_0(0)_R$}

The imaginary part ${\cal I}m\,f^{K^-p}_0(0)_R$ of the amplitude
$f^{K^-p}_0(0)_R$ is determined by inelastic channels. Near threshold
low--energy $K^-p$ interaction contains four inelastic channels
defined by strong low--energy interactions: (i) $K^-p \to
\Sigma^-\pi^+ $, (ii) $K^-p \to \Sigma^+\pi^-$, (iii) $K^-p \to
\Sigma^0\pi^0$ and (iv) $K^-p \to \Lambda^0\pi^0$. The amplitudes of
these channels we define as \cite{MR96}
\begin{eqnarray}\label{label4.6}
f(K^-p \to \Sigma^-\pi^+) &=&
\frac{1}{4\pi}\,\frac{\mu}{~m_{K^-}}\,\sqrt{\frac{m_{\Sigma^-}}{m_p}}
\Big[- \frac{g^2_1}{m_{\Lambda^0_1} - m_{K^-} - m_p} +
\frac{1}{6}\,\frac{g^2_2\,(1 + 3\alpha_2)}{m_{\Lambda^0_2} - m_{K^-} -
m_p}\nonumber\\ &-& \frac{1}{2}\,\frac{g^2_2\,\alpha_2\,(1 -
\alpha_2)}{m_{\Sigma^0_2} - m_{K^-} - m_p}\Big],\nonumber\\ f(K^-p \to
\Sigma^+\pi^-) &=& \frac{1}{4\pi}\,\frac{\mu}{~m_{K^-}}\,
\sqrt{\frac{m_{\Sigma^+}}{m_p}} \Big[- \frac{g^2_1}{m_{\Lambda^0_1} -
m_{K^-} - m_p} + \frac{1}{6}\,\frac{g^2_2(1 +
3\alpha_2)}{m_{\Lambda^0_2} - m_{K^-} - m_p}\nonumber\\ &+&
\frac{1}{2}\,\frac{g^2_2\,\alpha_2\,(1 - \alpha_2)}{m_{\Sigma^0_2} -
m_{K^-} - m_p}\Big],\nonumber\\ f(K^-p \to \Sigma^0\pi^0) &=&
\frac{1}{4\pi}\,\frac{\mu}{~m_{K^-}}\,\sqrt{\frac{m_{\Sigma^0}}{m_p}}
\Big[- \frac{g^2_1}{m_{\Lambda^0_1} - m_{K^-} - m_p} +
\frac{1}{6}\,\frac{g^2_2\,(1 + 3\alpha_2)}{m_{\Lambda^0_2} - m_{K^-} -
m_p}\Big],\nonumber\\ f(K^-p \to \Lambda^0\pi^0) &=&
\frac{1}{4\pi}\,\frac{\mu}{~m_{K^-}}\,\sqrt{\frac{m_{\Lambda^0}}{m_p}}
\Big[- \frac{1}{2}\,\frac{1}{\sqrt{3}}\,\frac{g^2_2\,(1 -
\alpha_2)}{m_{\Sigma^0_2} - m_{K^-} - m_p}\Big],
\end{eqnarray}
where $\alpha_2 = f_2/g_2$.

In order to check a consistency of our approach we suggest to use
experimental data on the cross sections for the inelastic reactions
$K^-p \to \Sigma^-\pi^+ $, $K^-p \to \Sigma^+\pi^-$, $K^-p \to
\Sigma^0\pi^0$ and $K^-p \to \Lambda^0\pi^0$ taken at threshold of the
$K^-p$ pair \cite{DT71,RN78}
\begin{eqnarray}\label{label4.7}
\gamma &=& \frac{\sigma(K^-p \to \Sigma^-\pi^+)}{\sigma(K^-p \to
\Sigma^+\pi^-)} = 2.360 \pm 0.040,\nonumber\\ R_c &=&
\frac{\sigma(K^-p \to \Sigma^-\pi^+) + \sigma(K^-p \to
\Sigma^+\pi^-)}{\sigma(K^-p \to \Sigma^-\pi^+) + \sigma(K^-p \to
\Sigma^+\pi^-) + \sigma(K^-p \to \Sigma^0\pi^0) + \sigma(K^-p \to
\Lambda^0\pi^0)} = \nonumber\\ &=&0.664 \pm 0.011,\nonumber\\ R_n &=&
\frac{\sigma(K^-p \to \Lambda^0\pi^0)}{\sigma(K^-p \to \Sigma^0\pi^0)
+ \sigma(K^-p \to \Lambda^0\pi^0)} = 0.189 \pm 0.015.
\end{eqnarray}
These data should place constraints on the input parameters of any
approach \cite{WW95}. In terms of the amplitudes of inelastic
reactions under consideration they read
\begin{eqnarray}\label{label4.8}
\gamma &=& \frac{|f(K^-p \to \Sigma^-\pi^+)|^2
k_{\Sigma^-\pi^+}}{|f(K^-p \to \Sigma^+\pi^-)|^2
k_{\Sigma^+\pi^-}},\nonumber\\ R_c &=& \Big(|f(K^-p \to
\Sigma^-\pi^+)|^2 k_{\Sigma^-\pi^+} + |f(K^-p \to \Sigma^+\pi^-)|^2
k_{\Sigma^+\pi^-}\Big)\nonumber\\ &\times&\Big(|f(K^-p \to
\Sigma^-\pi^+)|^2 k_{\Sigma^-\pi^+} + |f(K^-p \to \Sigma^+\pi^-)|^2
k_{\Sigma^+\pi^-}\nonumber\\ &+& |f(K^-p \to \Sigma^0\pi^0)|^2
k_{\Sigma^0\pi^0} + |f(K^-p \to \Lambda^0\pi^0)|^2
k_{\Lambda^0\pi^0}\Big)^{-1},\nonumber\\R_n &=& \frac{|f(K^-p \to
\Lambda^0\pi^0)|^2k_{\Lambda^0\pi^0}}{|f(K^-p \to \Sigma^0\pi^0)|^2
k_{\Sigma^0\pi^0} + |f(K^-p \to \Lambda^0\pi^0)|^2
k_{\Lambda^0\pi^0}},
\end{eqnarray}
where $k_{AB}$ with $A = \Sigma^{\pm},\Sigma^{0}, \Lambda^0$ and $B =
\pi^{\pm}, \pi^0$ is a relative momentum of the $AB$ pair, calculated
at threshold
\begin{eqnarray}\label{label4.9}
k_{AB}(s) = \frac{1}{2\sqrt{s}}\,\sqrt{(s - (m_A + m_B)^2)(s - (m_A -
m_B)^2)}.
\end{eqnarray}
At threshold $s = (m_{K^-} + m_p)^2$ and $k_{AB}((m_{K^-} + m_p)^2) =
k_{AB}$.

Expressing the amplitudes of inelastic channels with neutral particles
in the final states in terms of the amplitudes of the reactions with
charged particles in the final state we get
\begin{eqnarray}\label{label4.10}
f(K^-p \to \Sigma^0\pi^0) &=&
\frac{1}{2}\Big[\sqrt{\frac{m_{\Sigma^0}}{m_{\Sigma^-}}}\,f(K^-p \to
\Sigma^-\pi^+) + \sqrt{\frac{m_{\Sigma^0}}{m_{\Sigma^+}}}\,f(K^-p \to
\Sigma^+\pi^-)\Big],\nonumber\\ f(K^-p \to \Lambda^0\pi^0)
&=&\frac{1}{\alpha_2}
\frac{1}{2\sqrt{3}}\Big[\sqrt{\frac{m_{\Lambda^0}}{m_{\Sigma^-}}}
\,f(K^-p \to \Sigma^-\pi^+) -
\sqrt{\frac{m_{\Lambda^0}}{m_{\Sigma^+}}}\,f(K^-p \to
\Sigma^+\pi^-)\Big].\nonumber\\ &&
\end{eqnarray}
Combining the relations (\ref{label4.10}) and (\ref{label4.8}) we
express the amplitudes of inelastic channels $K^-p \to \Sigma^+\pi^-$,
$K^-p \to \Sigma^0\pi^0$ and $K^-p \to \Lambda^0\pi^0$ in terms of the
amplitude of the reaction $K^-p \to \Sigma^-\pi^+$. This gives
\begin{eqnarray}\label{label4.11}
f(K^-p \to \Sigma^+\pi^-) &=& f(K^-p\to \Sigma^-\pi^+)
\sqrt{\frac{1}{\gamma}\,\frac{k_{\Sigma^-\pi^+}}{k_{\Sigma^+\pi^-}}},
\nonumber\\ f(K^-p \to \Sigma^0\pi^0) &=& f(K^-p\to
\Sigma^-\pi^+)\frac{1}{2}\,\sqrt{\frac{m_{\Sigma^0}}{m_{\Sigma^-}}}
\,\Bigg(1 +
\sqrt{\frac{1}{\gamma}\,\frac{m_{\Sigma^-}}{m_{\Sigma^+}}\,
\frac{k_{\Sigma^-\pi^+}}{k_{\Sigma^+\pi^-}}}\,\Bigg),\nonumber\\
f(K^-p \to \Lambda^0\pi^0) &=&f(K^-p\to
\Sigma^-\pi^+)\,\sqrt{\frac{R_n}{1 -
R_n}\,\frac{k_{\Sigma^0\pi^0}}{k_{\Lambda^0\pi^0}}}\nonumber\\
&&\times\,\frac{1}{2}\,\sqrt{\frac{m_{\Lambda^0}}{m_{\Sigma^-}}}
\,\Bigg(1 + \sqrt{\frac{1}{\gamma}\,
\frac{m_{\Sigma^-}}{m_{\Sigma^+}}\,
\frac{k_{\Sigma^-\pi^+}}{k_{\Sigma^+\pi^-}}}\,\Bigg).
\end{eqnarray}
The parameter $\alpha_2$ is defined by
\begin{eqnarray}\label{label4.12}
\alpha_2 = - \sqrt{\frac{1 - R_n}{3
R_n}\,\frac{k_{\Lambda^0\pi^0}}{k_{\Sigma^0\pi^0}}}\;
\frac{\displaystyle 1 - \sqrt{\frac{1}{\gamma}\,
\frac{m_{\Sigma^-}}{m_{\Sigma^+}}\,
\frac{k_{\Sigma^-\pi^+}}{k_{\Sigma^+\pi^-}}}}{\displaystyle 1 +
\sqrt{\frac{1}{\gamma}\, \frac{m_{\Sigma^-}}{m_{\Sigma^+}}\,
\frac{k_{\Sigma^-\pi^+}}{k_{\Sigma^+\pi^-}}}}.
\end{eqnarray}
In our approach the parameter $R_c$ turns out to be dependent and
reads
\begin{eqnarray}\label{label4.13}
\hspace{-0.3in}R_c =\frac{1}{\displaystyle 1 +
\frac{1}{4}\frac{\gamma}{\gamma +
1}\,\frac{k_{\Sigma^0\pi^0}}{k_{\Sigma^-\pi^+}}\,
\Bigg(\frac{m_{\Sigma^0}}{m_{\Sigma^-}} + \frac{R_n}{1 -
R_n}\,\frac{m_{\Lambda^0}}{m_{\Sigma^-}} \Bigg)\Bigg(1 +
\sqrt{\frac{1}{\gamma}\, \frac{m_{\Sigma^-}}{m_{\Sigma^+}}\,
\frac{k_{\Sigma^-\pi^+}}{k_{\Sigma^+\pi^-}}}\Bigg)^2}.
\end{eqnarray}
Using the experimental values of $\gamma$, $R_n$ and masses of baryons
and mesons \cite{DG00} we get
\begin{eqnarray}\label{label4.14}
R_c &=& ~~\, 0.626 \pm 0.007,\nonumber\\ \alpha_2 &=& -\,0.314 \pm
0.026,
\end{eqnarray}
where uncertainties are caused by the experimental errors of the
parameters $\gamma$ and $R_n$. 

Comparing the theoretical prediction $R_c = 0.626 \pm 0.007$ with the
experimental value $R_c = 0.664 \pm 0.011$ in (\ref{label4.7}) we can
argue that our approach to the description of $K^-p$ scattering near
threshold is consistent with experimental data on the cross sections
for the inelastic reactions within an accuracy better than 6$\%$.

Hence, using the relations $\gamma$ and $R_c$ for the cross sections
for the inelastic reactions we can write down
\begin{eqnarray}\label{label4.15}
\sigma(K^-p \to {\rm all}) = \sum_X \sigma(K^-p \to X) =
\frac{1}{R_c}\,\Big(1 + \frac{1}{\gamma}\Big)\,\sigma(K^-p \to
\Sigma^- \pi^+),
\end{eqnarray}
where $X = \Sigma^-\pi^+, \Sigma^+\pi^-, \Sigma^0\pi^0$ and
$\Lambda^0\pi^0$.

Due to the optical theorem the relation (\ref{label4.15}) determines
the imaginary part of the amplitude $f^{K^-p}_0(0)_R$. It reads
\begin{eqnarray}\label{label4.16}
{\cal I}m\,f^{K^-p}_0(0)_R = \frac{1}{R_c}\,\Big(1 +
\frac{1}{\gamma}\Big)\,|f(K^-p \to \Sigma^-\pi^+)|^2
k_{\Sigma^-\pi^+}.
\end{eqnarray}
Since in our approach ${\cal I}m\,f^{K^-p}_0(0) = {\cal
I}m\,f^{K^-p}_0(0)_R$, the relation (\ref{label4.16}) allows to
determine the total width of kaonic hydrogen $\Gamma_{1s}$ in terms of
the partial width of the decay $A_{Kp} \to \Sigma^- + \pi^+$
\cite{WW95}
\begin{eqnarray}\label{label4.17}
\Gamma_{1s} &=& \frac{1}{R_c}\,\Big(1 +
\frac{1}{\gamma}\Big)\,\Gamma(A_{Kp}\to \Sigma^-\pi^+) =
842.248\,{\cal I}m\,f^{K^-p}_0(0) =\nonumber\\ &=&
842.248\,\frac{1}{R_c}\,\Big(1 + \frac{1}{\gamma}\Big)\,|f(K^-p \to
\Sigma^-\pi^+)|^2k_{\Sigma^-\pi^+}\;{\rm eV}.
\end{eqnarray}
For the calculation of the numerical value of $f(K^-p \to
\Sigma^-\pi^+)$ we have to determine the coupling constant $g_1$ and
$g_2$. They can be obtained fitting the total experimental widths of
the resonances $\Lambda(1405)$, $\Lambda(1800)$ and $\Sigma(1750)$
\cite{DG00}. We would like to notice that within an accuracy better
than 6$\%$ we can set $\alpha_2 = -1/3$ and neglect the contribution
of the $\Lambda(1800)$ resonance. Therefore, the constant $g_2$ we
define from the experimental data on the $\Sigma(1750)$ resonance
only.

We would like to emphasize that the experimental data on the masses
and total widths of the $\Lambda(1405)$ and $\Sigma(1750)$ resonances
are rather ambiguous. Below we use only recommended values for the
masses and total widths of these resonances \cite{DG00}.

\subsubsection{The $\Lambda(1405)$ resonance}

The recommended values for the mass and total width of the
$\Lambda(1405)$ resonance are equal to $m_{\Lambda^0_1} = 1406\,{\rm
MeV}$ and $\Gamma_{\Lambda^0_1} = 50\,{\rm MeV}$ \cite{DG00a,RD91}.

The total width of the $\Lambda(1405)$--resonance is defined by the
decays $\Lambda(1405) \to \Sigma + \pi$ \cite{DG00}. Due to the
effective Lagrangian (\ref{label4.3}) the total width of the
$\Lambda(1405)$ resonance $\Gamma_{\Lambda^0_1}$ reads
\begin{eqnarray}\label{label4.18}
\Gamma_{\Lambda^0_1} &=& \frac{g^2_1}{8\pi}\,\frac{(m_{\Lambda^0_1} +
m_{\Sigma^+})^2 - m^2_{\pi^-}}{m^2_{\Lambda^0_1}}\,k_{\Sigma^+\pi^-} +
\frac{g^2_1}{8\pi}\,\frac{(m_{\Lambda^0_1} + m_{\Sigma^-})^2 -
m^2_{\pi^+}}{m^2_{\Lambda^0_1}}\,k_{\Sigma^-\pi^+}\nonumber\\ &+&
\frac{g^2_1}{8\pi}\,\frac{(m_{\Lambda^0_1} + m_{\Sigma^0})^2 -
m^2_{\pi^0}}{m^2_{\Lambda^0_1}}\,k_{\Sigma^0\pi^0}.
\end{eqnarray}
Setting $\Gamma_{\Lambda^0_1} = 50\, {\rm MeV}$ and
using the experimental values for the masses of the $\Sigma$ hyperon
and $\pi$ meson \cite{DG00}, we obtain the value of the coupling
constant $g_1$: $g_1 = 0.907$. 

\subsubsection{The $\Sigma(1750)$ resonance}

The recommended values for the mass and total width of the
$\Sigma(1750)$ resonance are equal to $m_{\Sigma^0_2} = 1750\,{\rm
MeV}$ and $\Gamma_{\Sigma^0_2} = 90\,{\rm MeV}$
\cite{DG00b,VA96}. From the Lagrangian (\ref{label4.3}) we define the
total width of the $\Sigma(1750)$ resonance
\begin{eqnarray}\label{label4.19}
\Gamma_{\Sigma^0_2} &=& \frac{g^2_2}{72\pi}\,\frac{(m_{\Sigma^0_2} +
m_{\Sigma^+})^2 - m^2_{\pi^-}}{m^2_{\Sigma^0_2}}\,k_{\Sigma^+\pi^-} +
\frac{g^2_2}{72\pi}\,\frac{(m_{\Sigma^0_2} + m_{\Sigma^-})^2 -
m^2_{\pi^+}}{m^2_{\Sigma^0_2}}\,k_{\Sigma^-\pi^+}\nonumber\\ &+&
\frac{g^2_2}{24\pi}\,\frac{(m_{\Sigma^0_2} + m_{\Lambda^0})^2 -
m^2_{\pi^0}}{m^2_{\Sigma^0_2}}\,k_{\Lambda^0\pi^0}
+\frac{g^2_2}{24\pi}\,\frac{(m_{\Sigma^0_2} + m_{\Sigma^0})^2 -
m^2_{\eta}}{m^2_{\Sigma^0_2}}\,k_{\Sigma^0\eta}\nonumber\\
&+&\frac{g^2_2}{18\pi}\,\frac{(m_{\Sigma^0_2} + m_p)^2 -
m^2_{K^-}}{m^2_{\Sigma^0_2}}\,k_{p K^-} +
\frac{g^2_2}{18\pi}\,\frac{(m_{\Sigma^0_2} + m_n)^2 -
m^2_{\bar{K}^0}}{m^2_{\Sigma^0_2}}\,k_{n \bar{K}^0},
\end{eqnarray}
where we have used $\alpha_2 = - 1/3$. Setting $\Gamma_{\Sigma^0_2} =
90\,{\rm MeV}$ and using experimental values for the masses of baryons
and mesons we get $g_2 = 1.123$.

\subsubsection{Numerical values of $f(K^-p \to \Sigma^-\pi^+)$ and 
imaginary part of $f^{K^-p}_0(0)_R$}

Setting $\alpha_2 = -1/3$ in (\ref{label4.6}) and using the coupling
constant $g_1= 0.907$ and $g_2 = 1.123$, calculated above, we obtain
the numerical value of the amplitude $f(K^-p \to \Sigma^-\pi^+)$
\begin{eqnarray}\label{label4.20}
f(K^-p \to \Sigma^-\pi^+) &=&
\frac{1}{4\pi}\frac{\mu}{~m_{K^-}}\sqrt{\frac{m_{\Sigma^-}}{m_p}}
\Big[- \frac{g^2_1}{m_{\Lambda^0_1} - m_{K^-} - m_p}+
\frac{2}{9}\,\frac{g^2_2}{m_{\Sigma^0_2} - m_{K^-} -
m_p}\Big]=\nonumber\\ &=& (0.379\pm 0.023)\;{\rm fm}
\end{eqnarray}
Due to the relation (\ref{label4.16}) this gives the imaginary part of
the amplitude $f^{K^-p}_0(0)_R$ 
\begin{eqnarray}\label{label4.21}
{\cal I}m\,f^{K^-p}_0(0)_R = (0.269\pm 0.032)\;{\rm fm}.
\end{eqnarray}
According to this value and the relation ${\cal I}m\,f^{K^-p}_0(0) =
{\cal I}m\,f^{K^-p}_0(0)_R$ the total width $\Gamma_{1s}$ of kaonic
hydrogen in the ground state should be equal to
\begin{eqnarray}\label{label4.22}
\Gamma^{\rm th}_{1s} = 842.248\,{\cal I}m\,f^{K^-p}_0(0) = (227\pm
27)\,{\rm eV}.
\end{eqnarray}
This agrees well with recent experimental data by the DEAR
Collaboration $\Gamma^{\exp}_{1s} = (213\pm 138)\,{\rm eV}$
\cite{DEAR4}.

\subsection{Real part of $f^{K^-p}_0(0)_R$}

A knowledge of the numerical values of the coupling constants $g_1$,
$g_2$ and $\alpha_2$ allows to calculate the real part of the
amplitude $f^{K^-p}_0(0)_R$. In our approach it reads
\begin{eqnarray}\label{label4.23}
{\cal R}e\,f^{K^-p}_0(0)_R &=&\frac{1}{2}\,\Big({\cal
R}e\,f^{K^-p}_0(0)^{I = 0}_R + {\cal R}e\,f^{K^-p}_0(0)^{I = 1}_R\Big)
=\nonumber\\ &=&
\frac{1}{8\pi}\,\frac{\mu}{~~m_{K^-}}\,\Bigg[\frac{g^2_1}{\displaystyle
m_{\Lambda^0_1} - m_{K^-} - m_p} +
\frac{4}{9}\,\frac{g^2_2\,}{m_{\Sigma^0_2} - m_{K^-} -
m_p}\Bigg]=\nonumber\\ &=& (- 0.154 \pm 0.009)\;{\rm fm},
\end{eqnarray}
where we have set $\alpha_2 = -1/3$.

Now we proceed to the analysis of the contribution of a smooth elastic
background of low--energy elastic $K^-p$ scattering.

\section{Amplitude of low--energy $K^-p$ scattering. 
Elastic background }
\setcounter{equation}{0}

At the hadronic level a smooth elastic background $A^{K^-p}_B$ we
define as
\begin{eqnarray}\label{label5.1}
A^{K^-p}_B = A^{K^-p}_s + A^{K^-p}_t + A^{K^-p}_u,
\end{eqnarray}
where $A^{K^-p}_s$, $A^{K^-p}_t$ and $A^{K^-p}_u$ are the
contributions of the $s$, $t$ and $u$ channels of low--energy elastic 
$K^-p$ scattering, respectively.

For the calculation of the r.h.s. of (\ref{label5.1}) we assume the
following contributions
\begin{eqnarray}\label{label5.2}
A^{K^-p}_B = A^{K^-p}_{\rm CA} + A^{K^-p}_{\bar{K}K},
\end{eqnarray}
where (i) $A^{K^-p}_{\rm CA}$ is defined by the current algebra
\cite{SW66}--\cite{EK98}, accounting for all low--energy interactions
which can be described by Effective Chiral Lagrangians \cite{SG69}. In
the general form this contribution has been calculated in
\cite{BC79,EK98}; (ii) $A^{K^-p}_{\bar{K}K}$ is the contribution of
the four--quark intermediate states $qq\bar{q}\bar{q}$ (or $\bar{K}K$
molecule) such as the scalar mesons $a_0(980)$, $f_0(980)$ and so on
\cite{RJ77}--\cite{SK03} (see also \cite{AI00}) going beyond the scope
of Effective Chiral Lagrangians. As has been recently found by the
KLOE Collaboration (DAPHNE), measuring the radiative decays of the
vector $\phi(1020)$--meson, $\phi(1020) \to a_0(980)\gamma$ and
$\phi(1020) \to f_0(980)$, that the quark structure of the scalar
mesons $a_0(980)$ and $f_0(980)$ differs substantially from $q\bar{q}$
\cite{DAPHNE}.

\subsection{Calculation of $A^{K^-p}_{\rm CA}$}

The current algebra contribution to the parameter $A^{K^-p}_B$ we
denote as
\begin{eqnarray}\label{label5.3}
A^{K^-p}_{\rm CA} = \frac{1}{2}\,(A^0_0 + A^1_0),
\end{eqnarray}
where $A^0_0$ and $A^1_0$ describe the contribution of $K^-p$
scattering in the states with isospin $I = 0$ and $I = 1$. Using the
results obtained in \cite{BC79,EK98} we get
\begin{eqnarray}\label{label5.4}
A^0_0 &=& \frac{3}{8\pi}\,\frac{\mu}{F^2_K},\nonumber\\
A^1_0 &=& \frac{1}{8\pi}\,\frac{\mu}{F^2_K},
\end{eqnarray}
where $F_K = 112.996\,{\rm MeV}$ is the PCAC constant of the $K^{\pm}$
meson \cite{DG00}. This gives
\begin{eqnarray}\label{label5.5}
A^{K^-p}_{\rm CA} = \frac{1}{4\pi}\,\frac{\mu}{F^2_K} = 0.398\,{\rm
fm}.
\end{eqnarray}
The value (\ref{label5.5}) is caused by the contributions of the $s$,
$t$ and $u$ channels of low--energy elastic $K^-p$ scattering, which
can be described by Effective Chiral Lagrangians \cite{SG69}. The
result (\ref{label5.5}) is obtained at leading order in Chiral
perturbation theory \cite{JG99,JG83} (see also \cite{ML02}). According
to Chiral perturbation theory \cite{JG99,JG83} the accuracy of the
value, given by (\ref{label5.5}), is of order of $O(m^2_{K^-}/16\pi^2
F^2_K) = O(12\%)$. This coincides with an accuracy of the current
algebra approach \cite{SA68,HP73}.

\subsection{Calculation of four--quark contribution 
$A^{K^-p}_{\bar{K}K}$}

Four--quark states (or $\bar{K}K$ molecule) such as the scalar mesons
$a_0(980)$ and $f_0(980)$ can give a contribution only to the
$t$--channel of low--energy elastic $K^-p$ scattering defined by the
reaction $K^- + K^+ \to p + \bar{p}$. Since the four--quark states
$a_0(980)$ and $f_0(980)$ cannot be described by Effective Chiral
Lagrangians \cite{SG69}, the contribution of these states do not enter
to $A^{K^-p}_{\rm CA}$.

According to Jaffe \cite{RJ77}, the scalar mesons $a_0(980)$ and
$f_0(980)$ belong to an $SU(3)_{\rm flavour}$ nonet and the scalar
meson $f_0(980)$ decouples from the $\pi \pi$ state. Following Jaffe
\cite{RJ77}, the $SU(3)_{\rm flavour}$ invariant interaction of the
nonet of four--quark scalar mesons with two nonents of pseudoscalar
light mesons, having a $q\bar{q}$ quark structure, can be written as
\begin{eqnarray}\label{label5.6}
{\cal L}_{SPP}(x) = \sqrt{2}\,g_0\,{\rm tr}\{PPM\} =
\sqrt{2}\,g_0\,P^b_aP^a_cM^c_b.
\end{eqnarray}
where $P$ and $M$ are nonets of pseudoscalar light $q\bar{q}$ mesons
and scalar $qq\bar{q}\bar{q}$ mesons, respectively,
\begin{eqnarray}\label{label5.7}
P^b_a &=& \left(\begin{array}{llcl} {\displaystyle
\frac{\pi^0}{\sqrt{2}} + \frac{\eta_0}{\sqrt{2}}} &
\hspace{0.3in}\pi^+ & K^+ \\ \hspace{0.3in}\pi^- &{\displaystyle -
\frac{\pi^0}{\sqrt{2}} + \frac{\eta_0}{\sqrt{2}}} & K^0 \\
\hspace{0.15in}- K^- & \hspace{0.3in}\bar{K}^0 & \eta_s \\
\end{array}\right),\nonumber\\
 M^b_a &=& \left(\begin{array}{llcl} {\displaystyle
\frac{a^0_0}{\sqrt{2}} - \frac{\varepsilon}{2}} &
\hspace{0.3in}a^+_0 & \kappa^+ \\ \hspace{0.3in}a^-_0 &{\displaystyle -
\frac{a^0_0}{\sqrt{2}} - \frac{\varepsilon}{2}} & \kappa^0 \\
\hspace{0.15in} - \kappa^- & \hspace{0.3in}\bar{\kappa}^0 &
{\displaystyle - \frac{f_0}{\sqrt{2}} + \frac{\varepsilon}{2}} \\
\end{array}\right),
\end{eqnarray}
where $\eta_0$ and $\eta_s$ are pseudoscalar states with quark
structure $\eta_0 = (u\bar{u} + d\bar{d})/\sqrt{2}$ and $\eta_s =
s\bar{s}$ \cite{RJ77}. Then, $\vec{a}_0 = (a^+_0, a^0_0, a^-_0) =
(s\bar{s}u\bar{d}, s\bar{s}(u\bar{u} - d\bar{d})/\sqrt{2},
d\bar{u}s\bar{s})$ is the isotriplet of $a_0(980)$ mesons, $\kappa =
(\kappa^+,\kappa^0) = (u\bar{s}d\bar{d},d\bar{s}u\bar{u})$ and
$\bar{\kappa} = (\bar{\kappa}^0, - \kappa^-) = (s\bar{d}u\bar{u}, -
s\bar{u}d\bar{d})$ are doublets of strange scalar four--quark states,
$f_0 = s\bar{s}(u\bar{u} - d\bar{d})/\sqrt{2}$ is the $f_0(980)$ meson
and $\varepsilon$ is the isoscalar scalar $\varepsilon(700)$ meson
with $\varepsilon = u\bar{d}d\bar{u}$ quark structure and mass
$m_{\varepsilon} = 700\,{\rm MeV}$ \cite{RJ77}. The nonet $M$ is
constructed in such a way that the $f_0(980)$ meson decouples from the
$\pi \pi$ states, whereas the $\varepsilon(700)$ meson couples to the
$\pi \pi$ states but decouples from the $\bar{K}K$ states. This
implies that the $\varepsilon(700)$ meson does not contribute to the
amplitude of $K^-p$ scattering.

The interactions of the scalar mesons $a_0(980)$ and $f_0(980)$ with
the $K^-$--meson can be written as
\begin{eqnarray}\label{label5.8}
{\cal L}_{SKK}(x) = g_0\,[- a^0_0(x) + f_0(x)]\,K^+(x)K^-(x),
\end{eqnarray}
where $a^0_0(x)$, $f_0(x)$ and $K^{\pm}(x)$ are interpolating fields
of the $a^0_0(980)$, $f_0(980)$ and $K^{\pm}$ mesons. 

For a numerical calculation we use the value $g_0 = g_{a_0K^+K^-} =
g_{f_0K^+K^-} = 2.746\,{\rm GeV}$, obtained within the
$\bar{K}K$ molecule model of the scalar mesons $a_0(980)$ and
$f_0(980)$ \cite{FC93} (see also \cite{NA80} and \cite{JW82}). In this
model the scalar mesons $a_0(980)$ and $f_0(980)$ couple only to the
$\bar{K}K$ states\,\footnote{The scalar mesons $a_0(980)$ couples also
to the $\pi \eta$ states, where $\eta$ is the well--known $\eta(550)$
pseudoscalar meson \cite{DG00}.} and decouple from the $\pi \pi$
states. 

The interaction of the nonet of four--quark scalar mesons $M$ with
octets of light baryons we define as
\begin{eqnarray}\label{label5.9}
{\cal L}_{SBB}(x) &=& \sqrt{2} g_D\,{\rm tr}\{\{\bar{B},B\}M\} +
\sqrt{2} g_F\,{\rm tr}\{[\bar{B},B]M\} = \nonumber\\ &=&\sqrt{2}(g_D +
g_F)\bar{B}^b_aB^a_cM^c_b + \sqrt{2}(g_D - g_F)\bar{B}^b_aB^c_bM^a_c,
\end{eqnarray}
where $B$ and $\bar{B}$ are octets of light baryons (see
(\ref{label4.2}))
\begin{eqnarray}\label{label5.10}
 \bar{B}^b_a = \left(\begin{array}{llcl} {\displaystyle
\frac{\bar{\Sigma}^0}{\sqrt{2}} + \frac{\bar{\Lambda}^0}{\sqrt{6}}} &
\hspace{0.3in}\bar{\Sigma}^- & - \bar{\Xi}^- \\
\hspace{0.3in}\bar{\Sigma}^+ &{\displaystyle -
\frac{\bar{\Sigma}^0}{\sqrt{2}} + \frac{\bar{\Lambda}^0}{\sqrt{6}}} &
\bar{\Xi}^0 \\
\hspace{0.3in}\bar{p}& \hspace{0.3in} \bar{n} & {\displaystyle
-\frac{2}{\sqrt{6}}\bar{\Lambda}^0} \\
\end{array}\right)
\end{eqnarray}
and $g_D$ and $g_F$ are the coupling constants of the symmetric and
antisymmetric $SBB$ interactions \cite{JS63}.

The effective Lagrangian of the $SNN$ interaction reads
\begin{eqnarray}\label{label5.11}
\hspace{-0.3in}&&{\cal L}_{SBB}(x) = (g_D +
g_F)\,[\sqrt{2}\,\bar{p}(x)n(x)a^+_0(x) +
\sqrt{2}\,\bar{n}(x)p(x)a^-_0(x)\nonumber\\ \hspace{-0.3in}&&+
(\bar{p}(x)p(x) - \bar{n}(x)n(x))a^0_0(x) - (1 -
2\alpha_S)\,(\bar{p}(x)p(x) + \bar{n}(x)n(x))f_0(x)\nonumber\\
\hspace{-0.3in}&&- \sqrt{2}\alpha_S\,(\bar{p}(x)p(x) +
\bar{n}(x)n(x))\varepsilon(x)] + \ldots,
\end{eqnarray}
where $\varepsilon(x)$, $p(x)$ and $n(x)$ are the interpolating fields
of the $\varepsilon(700)$ meson, the proton and the neutron.  The
parameter $\alpha_S$ is given by $\alpha_S = g_F/(g_D + g_F)$
\cite{JS63}.

In order to suppress the contribution of the four--quark state
$\varepsilon(700)$ to the S--wave scattering lengths of $\pi N$
scattering we have to set $\alpha_S = 0$ or $g_F = 0$. As a result the
four--quark state $\varepsilon(700)$ decouples from nucleons. This
gives
\begin{eqnarray}\label{label5.12}
\hspace{-0.3in}&&{\cal L}_{SBB}(x) =
g_D\,[\sqrt{2}\,\bar{p}(x)n(x)a^+_0(x) +
\sqrt{2}\,\bar{n}(x)p(x)a^-_0(x)\nonumber\\ \hspace{-0.3in}&&+
(\bar{p}(x)p(x) - \bar{n}(x)n(x))a^0_0(x) - (\bar{p}(x)p(x) +
\bar{n}(x)n(x))f_0(x)] + \ldots\,.
\end{eqnarray}
At threshold of the reaction $K^- + p \to K^- + p$ the contribution of
the four--quark states $a_0(980)$ and $f_0(980)$ we define as
\begin{eqnarray}\label{label5.13}
A^{K^-p}_{\bar{K}K} = \frac{M(K^- p\to K^- p)_{a_0 + f_0}}{8\pi
(m_{K^-} + m_p)} = - \frac{g_D}{2\pi}
\,\frac{g_0}{~m^2_{a_0}}\,\frac{\!\mu}{~m_{K^-}},
\end{eqnarray}
where we have set $m_{a_0} = m_{f_0} = 980\,{\rm MeV}$ \cite{DG00}.

The coupling constant $g_D$ is not known \cite{Gora}. For a further
calculation of $A^{K^-p}_{\bar{K}K}$ we can set \cite{AI90}
\begin{eqnarray}\label{label5.14}
g_D = \frac{g_{\pi NN}}{g_A}\,\xi,
\end{eqnarray}
where $g_{\pi NN} = 13.21$ \cite{PSI2} and $g_A = 1.267$ are the $\pi
NN$ coupling constant and the renormalization constant of the
axial--vector coupling due to strong interactions, and $\xi$ is a
parameter, which we estimate below.

Using (\ref{label5.14}) the contribution of the $a_0(980)$ and
$f_0(980)$ scalar mesons can be written as
\begin{eqnarray}\label{label5.15}
\hspace{-0.3in}A^{K^-p}_{\bar{K}K} = \frac{M(K^- p \to K^- p)_{a_0 +
f_0}}{8\pi (m_{K^-} + m_p)} = -\,\xi\,\frac{1}{2\pi}\frac{g_{\pi
NN}}{~g_A} \,\frac{g_0}{~m^2_{a_0}}\,\frac{\!\mu}{~m_{K^-}} = -
0.614\,\xi\,{\rm fm}.
\end{eqnarray}
The parameter $A^{K^-p}_B$ is equal to
\begin{eqnarray}\label{label5.16}
A^{K^-p}_B = 0.398 - 0.614\,\xi\,{\rm fm}.
\end{eqnarray}
In order estimate the value of the parameter $\xi$ we suggest to
calculate the parameter $A^{K^-p}_B$ within the Effective quark model
with chiral $U(3)\times U(3)$ symmetry \cite{AI99}--\cite{AI92}.

\subsection{$A^{K^-p}_B$ in effective quark model 
with chiral $U(3)\times U(3)$ symmetry}

Following the principle of the quark--hadron duality \cite{VZ79} we
assume that the contribution of the smooth elastic background of
low--energy elastic $K^-p$ scattering can be fully fitted by the
lowest quark box--diagram depicted in Fig.1, calculated with the
Effective quark model with chiral $U(3)\times U(3)$ symmetry
\cite{AI99}--\cite{AI92}.
\begin{figure}
  \centering
  \includegraphics[width=0.35\textwidth]{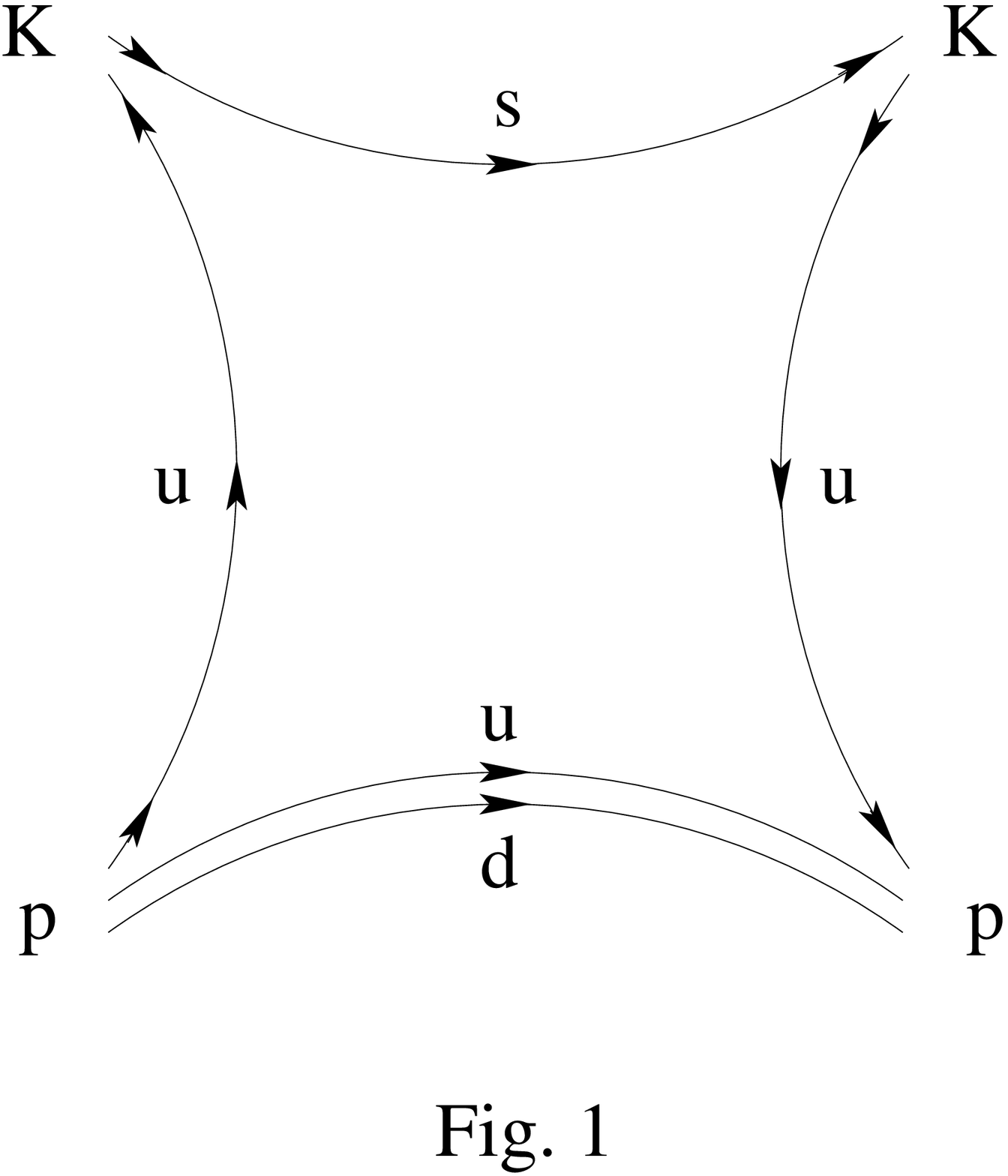}
\caption{The quark diagram describing a smooth elastic background of
low--energy elastic $K^-p$ scattering in the Effective quark model
with chiral $U(3)\times U(3)$ symmetry.}
\label{figpot}
\end{figure}

Using the reduction technique \cite{IZ80} the amplitude of elastic
low--energy $K^-p$ scattering we define as
\begin{eqnarray}\label{label5.17}
\hspace{-0.3in}&&(2\pi)^4 i\,\delta^{(4)}(q\,' + p\,' - q - p)\,M(K^-
p \to K^- p) =\nonumber\\
\hspace{-0.3in}&& = \lim_{p\,'^{\,2},\, p^2 \to m^2_p,\, q\,'^{\,2},
\,q^2 \to m^2_{K^-}}\int d^4x_1 d^4x_2 d^4x_3 d^4x_4\,e^{\textstyle
i\,q\,' \cdot x_1 + ip\,'\cdot x_2 - ip\cdot x_3 - i q\cdot
x_4}\nonumber\\
\hspace{-0.3in}&&\times\,(\Box_1 + m^2_{K^-})(\Box_4 +
m^2_{K^-})\,\bar{u}(p\,',\sigma\,'\,)\,\overrightarrow{(i\gamma_{\nu}
\,\partial^{\nu}_2 - m_p)}\langle 0|{\rm T}(K^-(x_1) p(x_2)
\bar{p}(x_3) K^+(x_4))|0\rangle\nonumber\\
\hspace{-0.3in}&&\times\,\overleftarrow{(-
i\gamma_{\mu}\,\partial^{\mu}_3 - m_p)}\,u(p,\sigma),
\end{eqnarray}
where $p(x)$ and $u(p,\sigma)$ are the interpolating field operator
and the Dirac bispinor of the proton, and $K^{\pm}(x)$ are the
interpolating fields of the $K^{\mp}$--mesons.

In order to describe the r.h.s. of Eq.(\ref{label5.17}) at the quark
level we follow \cite{AI99} and use the equations of motion
\begin{eqnarray}\label{label5.18}
\overrightarrow{(i\gamma_{\nu}\,\partial^{\nu}_2 - m_p)}\,p(x_1) &=&
\frac{g_{\rm B}}{\sqrt{2}}\,\eta_p(x_2),\nonumber\\
\bar{p}(x_3)\overleftarrow{(- i\gamma_{\mu}\partial^{\mu}_3 - m_p)}
&=& \frac{g_{\rm B}}{\sqrt{2}}\,\bar{\eta}_p(x_3),
\end{eqnarray}
where $\eta_p(x_2)$ and $\bar{\eta}_p(x_3)$ are the three--quark
current densities \cite{AI99}
\begin{eqnarray}\label{label5.19}
\eta_p(x_2) &=& -
\,\varepsilon^{ijk}\,[\bar{u^c}_i(x_2)
\gamma^{\mu}u_j(x_2)]\gamma_{\mu}\gamma^5 d_k(x_2),\nonumber\\ 
\bar{\eta}_p(x_3) &=&+\,\varepsilon^{ijk}\,\bar{d}_i(x_3)\gamma^{\mu}
\gamma^5[\bar{u}_j(x_3)\gamma_{\mu}u^c_k(x_3)]
\end{eqnarray}
where $i,j$ and $k$ are colour indices and $\bar{\psi^{\,c}}(x) =
\psi(x)^T C$ and $C = - C^T = - C^{\dagger} = - C^{-1} $ is the charge
conjugate matrix, $T$ denotes transposition, and $g_{\rm B}$ is the
phenomenological coupling constant of the low--lying baryon octet
$B_8(x)$ coupled to the three--quark current densities \cite{AI99}
\begin{eqnarray}\label{label5.20}
\hspace{-0.5in}{\cal L}^{(\rm B)}_{\rm int}(x) = \frac{g_{\rm
B}}{\sqrt{2}}\,\bar{B}_8(x)\eta_8(x) + {\rm h.c.}
\end{eqnarray}
The coupling constant $g_{\rm B}$ is equal to $g_{\rm B} = 1.34\times
10^{-4}\,{\rm MeV}^{-2}$ \cite{AI99}.

For the interpolating field operators of the $K^{\pm}$--mesons we use
the following equations of motion \cite{AI99}
\begin{eqnarray}\label{label5.21}
(\Box_1 + m^2_{K^-})K^-(x_1) &=&
\frac{g_K}{\sqrt{2}}\,\bar{u}(x_1)i\gamma^5 s(x_1),\nonumber\\ (\Box_4
+ m^2_{K^-})K^+(x_4) &=& \frac{g_K}{\sqrt{2}}\,\bar{s}(x_4)i\gamma^5
u(x_4),
\end{eqnarray}
where $g_K = (m + m_s)/\sqrt{2}F_K$, $m = 330 \,{\rm MeV}$ and $m_s =
465\,{\rm MeV}$ are the masses of the constituent $u$, $d$ and $s$
quarks, respectively \cite{AI99,AI92} (see also \cite{AI80}).

The amplitude of low--energy elastic  $K^-p$ scattering is defined by
\begin{eqnarray}\label{label5.22}
\hspace{-0.3in}&&M(K^- p \to K^- p) = -\,i\,\frac{1}{4}\,g^2_{\rm
B}\,g^2_K\int d^4x_1 d^4x_2 d^4x_3\,e^{\textstyle i\,q\,' \cdot x_1 +
ip\,'\cdot x_2 - ip\cdot x_3}\nonumber\\
\hspace{-0.3in}&&\times\,\bar{u}(p\,',\sigma\,'\,)\langle 0|{\rm
T}(\bar{u}(x_1)i\gamma^5 s(x_1)\eta_p(x_2) \bar{\eta}_p(x_3)
\bar{s}(0)i\gamma^5 u(0))|0\rangle\,u(p,\sigma).
\end{eqnarray}
where the external momenta $q\,'$, $p\,'$, $q$ and $p$ should be kept
on mass shell $q\,'^{\,2} = q^2 = m^2_{K^-}$ and $p\,'^{\,2} = p^2 =
m^2_p$.

In the Appendix we have carried out the calculation of the amplitude
(\ref{label5.22}) at threshold. The parameter $A^{K^-p}_B$ is equal to
(see ({\rm A}.9))
\begin{eqnarray}\label{label5.23}
A^{K^-p}_B = \frac{M(K^- p \to K^- p)}{8\pi (m_{K^-} + m_p)} = -
0.328 \pm 0.033\,{\rm fm}.
\end{eqnarray}
This allows to estimate the value of the parameter $\xi$
(\ref{label5.14}). Equating (\ref{label5.16}) to (\ref{label5.23}) we
get $\xi = 1.2\pm 0.1$.

\subsection{S--wave scattering length $a^{K^-p}_0$ and shift 
$\epsilon^{\rm th}_{1s}$}

Using the value of the parameter $A^{K^-p}_B$, describing the
contribution of the smooth elastic background of low--energy elastic 
$K^-p$ scattering, we obtain the S--wave scattering length
$a^{K^-p}_0$
\begin{eqnarray}\label{label5.24}
a^{K^-p}_0 = (- 0.328\pm 0.033) + (- 0.154 \pm 0.009) = (- 0.482 \pm
0.034)\,{\rm fm}.
\end{eqnarray}
This results in the shift of the energy level of the ground state of
kaonic hydrogen
\begin{eqnarray}\label{label5.25}
\epsilon^{\rm th}_{1s} = -\,421.124\,a^{K^-p}_0 = 203 \pm 15\;{\rm eV}.
\end{eqnarray}
The theoretical value fits well the preliminary experimental data
$\epsilon^{\exp}_{1s} = (183 \pm 62)\,{\rm eV}$ by the DEAR
Collaboration \cite{DEAR4}.

\section{Electromagnetic decay channels}
\setcounter{equation}{0}

It is well--known \cite{PSI2} that in the case of the energy level
displacement of the ground state of pionic hydrogen the
electromagnetic channel $A_{\pi p} \to n + \gamma$ defines 64$\%$ of
the experimental value of the width $\Gamma_{1s} = (0.868 \pm
0.056)\,{\rm eV}$. The width of the energy level of the ground state
of pionic hydrogen can be written as
\begin{eqnarray}\label{label6.1}
\Gamma_{1s}&=&
\frac{8\pi}{9}\,\frac{p^*}{\mu}\,(a^{1/2}_0 -
a^{3/2}_0)^2\,|\Psi_{1s}(0)|^2\,\Big(1 + \frac{1}{P}\Big),
\end{eqnarray}
where $\mu = m_{\pi^-}m_p/(m_{\pi^-} + m_p) = 121.497\,{\rm MeV}$ is
the reduced mass of the $\pi^-p$ system for $m_{\pi^-} = 139.570\,{\rm
MeV}$ and $m_p = 938.272\,{\rm MeV}$, $p^*$ is the relative momentum
equal to
\begin{eqnarray}\label{labe6.2}
p^* = \frac{m_p + m_{\pi^-}}{2}\sqrt{\Big[1 - \Big(\frac{m_n +
m_{\pi^0}}{m_p + m_{\pi^-}}\Big)^2\Big]\Big[1 - \Big(\frac{m_n -
m_{\pi^0}}{m_p + m_{\pi^-}}\Big)^2\Big]} = 28.040\,{\rm MeV},
\end{eqnarray}
$\Psi_{1s}(0) = 1/\sqrt{\pi a^3_B}$ is the wave function of the ground
state of pionic hydrogen at the origin, and $a^{1/2}_0$ and
$a^{3/2}_0$ are the S--wave scattering lengths of $\pi N$ scattering
with isospin $I = 1/2$ and $I = 3/2$. The experimental values
$a^{1/2}_0 = 0.1788 \pm 0.0043\,m^{-1}_{\pi^-}$ and $a^{3/2}_0 = -
0.0927 \pm 0.0085\,m^{-1}_{\pi^-}$, obtained by the PSI Collaboration
\cite{PSI2}, give $a^{1/2}_0 - a^{3/2}_0 = 0.2715 \pm
0.0095\,m^{-1}_{\pi^-}$. Then, $P$ is the Panofsky ratio defined by
\cite{JS77}
\begin{eqnarray}\label{label6.3}
\frac{1}{P} = \frac{\Gamma(A_{\pi p} \to n \gamma)}{\Gamma(A_{\pi p}
\to n \pi^0)} = 0.647 \pm 0.004,
\end{eqnarray}
where we have adduced the experimental value of $1/P$ obtained in
\cite{JS77}. 

In the case of kaonic hydrogen there are two electromagnetic decay
channels $A_{Kp} \to \Lambda^0 + \gamma$ and $A_{Kp} \to \Sigma^0 +
\gamma$, which are related to the reactions $K^- + p \to \Lambda^0 +
\gamma$ and $K^- + p \to \Sigma^0 + \gamma$. Therefore, the total
width of the energy level of the ground state of kaonic hydrogen can
be written as \cite{JG03}
\begin{eqnarray}\label{label6.4}
\Gamma_{1s} = \frac{4\pi}{\mu}\,{\cal I}m\,f_{K^-p
}(0)\,|\Psi_{1s}(0)|^2\,(1 + X),
\end{eqnarray}
where $X$, the inverse Panofsky ratio for kaonic hydrogen, is defined
by \cite{JG03}
\begin{eqnarray}\label{label6.5}
X = \frac{\Gamma(A_{Kp} \to \Lambda^0\gamma) + \Gamma(A_{Kp} \to
\Sigma^0\gamma)}{\Gamma_{1s}}.
\end{eqnarray}
Below we give a theoretical analysis and numerical estimate of the
value of $X$.

First, we consider the decay of pionic hydrogen $A_{\pi p} \to n +
\gamma$, then we extend the developed technique and methodics to the
decays of kaonic hydrogen $A_{Kp} \to \Lambda^0 + \gamma$ and $A_{Kp}
\to \Sigma^0 + \gamma$.

\subsection{Radiative decay of pionic hydrogen}

The amplitude of the decay $A_{\pi p} \to n + \gamma$ we define as
\cite{IV1,IV2,SA68,HP73}
\begin{eqnarray}\label{label6.6}
M(A_{\pi p} \to n \gamma) = \sqrt{\frac{1}{2\mu}}\int
\frac{d^3k}{(2\pi)^3}\,\sqrt{\frac{m_{\pi^-}m_p}{E_{\pi^-}(\vec{k}\,)
E_p(\vec{k}\,)}}\,\Phi_{1s}(\vec{k}\,)\,M(\pi^-(\vec{k}\,)
p(-\vec{k}\,) \to n \gamma),
\end{eqnarray}
where $\mu = m_{\pi^-}m_p/(m_{\pi^-} + m_p) = 121.497\,{\rm MeV}$ is
the reduced mass of the $\pi^-p$ system and $\Phi_{1s}(\vec{k}\,)$ is
the wave function of the ground state of pionic hydrogen in the
momentum representation.  

The amplitude $M(\pi^-(\vec{k}\,) p(-\vec{k}\,) \to n \gamma)$ of the
reaction $\pi^- + p \to n + \gamma$ is determined by \cite{SA68,HP73}
\begin{eqnarray}\label{label6.7}
M(\pi^-(\vec{k}\,) p(-\vec{k}\,) \to n \gamma) =
\sqrt{4\pi}\,e\,\langle n(-\vec{q},\sigma)|J^{\rm e\ell
m}_{\mu}(0)|\pi^-(\vec{k}\,) p(-
\vec{k},\sigma_p)\rangle\,e^{\mu}(\vec{q},\lambda),
\end{eqnarray}
where $J^{\rm e\ell m}_{\mu}(0)$ is the electromagnetic hadronic
current \cite{SA68,HP73}
\begin{eqnarray}\label{label6.8}
J^{\rm e\ell m}_{\mu}(0) = J^3_{\mu}(0) +
\frac{1}{\sqrt{3}}\,J^8_{\mu}(0).
\end{eqnarray}
Here, $J^3_{\mu}(0)$ is the third component of the isotopic vector and
$J^8_{\mu}(0)$, the isospin singlet, is the eighth component of the
$SU(3)_{\rm flavour}$ octet; $e^{\mu}(\vec{q},\lambda)$ is the
polarization vector of the emitted photon.

Using the reduction technique \cite{IZ80} for the $\pi^-$--meson we
reduce the matrix element of the electromagnetic hadronic current
(\ref{label6.6}) to the form
\begin{eqnarray}\label{label6.9}
\hspace{-0.3in}\langle n(-\vec{q},\sigma)|J^{\rm e\ell
m}_{\mu}(0)|\pi^-(\vec{k}\,) p(-\vec{k},\sigma_p)\rangle &=&
\lim_{k^2_{\pi^-}\to m^2_{\pi^-}}i\int
d^4x\,e^{\textstyle\,-ik_{\pi^-}\cdot x}\,(\Box_x +
m^2_{\pi^-})\nonumber\\ \hspace{-0.3in}&\times&\langle
n(-\vec{q},\sigma)| {\rm T}(J^{\rm e\ell m}_{\mu}(0)
\pi^{-\dagger}(x))| p(-\vec{k},\sigma_p)\rangle,
\end{eqnarray}
where $k_{\pi^-} = (\sqrt{\vec{k}^{\,2} + m^2_{\pi^-}}, \vec{k}\,)$.
According to the PCAC hypothesis \cite{SA68,HP73} the interpolating
fields of the $\pi$ mesons are related to the divergences of the
axial--vector currents. For the $\pi^-$--meson field we get
\begin{eqnarray}\label{label6.10}
\pi^{-\dagger}(x) = \frac{1}{\sqrt{2}}\,\frac{1}{m^2_{\pi}F_{\pi}}\,
\partial^{\nu}J^{1-i2}_{5\nu}(x),
\end{eqnarray}
where $F_{\pi} = 92.419\,{\rm MeV}$ is the PCAC constant and
$J^{1-i2}_{5\nu}(x) = J^1_{5\nu}(x) - iJ^2_{5\nu}(x)$ is the hadronic
axial--vector current \cite{SA68,HP73}.

In the soft--pion limit \cite{SA68,HP73} the r.h.s. of
(\ref{label6.9}) can be rewritten as\,\footnote{The soft--pion limit
as well as the soft--kaon limit should be understood as ChPT at
leading order in chiral expansions \cite{JG99,JG83}.}
\begin{eqnarray}\label{label6.11}
&&\langle n(-\vec{q},\sigma)|J^{\rm e\ell
m}_{\mu}(0)|\pi^-(\vec{k}\,) p(-\vec{k},\sigma_p)\rangle =\nonumber\\
&&= \frac{i}{\sqrt{2}F_{\pi}} \int d^4x\,\langle
n(-\vec{q},\sigma)| {\rm T}(J^{\rm e\ell
m}_{\mu}(0)\partial^{\nu}J^{1-i2}_{5\nu}(x))|
p(\vec{0},\sigma_p)\rangle.
\end{eqnarray}
Integrating by parts we arrive at the expression \cite{SA68,HP73}
\begin{eqnarray}\label{label6.12}
&&\langle n(-\vec{q},\sigma)|J^{\rm e\ell
m}_{\mu}(0)|\pi^-(\vec{k}\,) p(-\vec{k},\sigma_p)\rangle =\nonumber\\
&&= \frac{i}{\sqrt{2}F_{\pi}}\langle n(-\vec{q},\sigma)|[J^{\rm
e\ell m}_{\mu}(0), Q^{1-i2}_5(0)]| p(\vec{0},\sigma_p)\rangle,
\end{eqnarray}
where $Q^{1-i2}_{5\nu}(0)$ is the axial--vector charge operator
\begin{eqnarray}\label{label6.13}
Q^{1-i2}_5(0) = \int d^3x\,J^{1-i2}_{5 0}(0,\vec{x}\,).
\end{eqnarray}
Using Gell--Mann's current algebra \cite{SA68,HP73} we get
\begin{eqnarray}\label{label6.14}
\langle n(-\vec{q},\sigma)|J^{\rm e\ell m}_{\mu}(0)|\pi^-(\vec{k}\,)
p(-\vec{k},\sigma_p)\rangle = -\,\frac{i}{\sqrt{2}F_{\pi}}\langle
n(-\vec{q},\sigma)|J^{1-i2}_{5\mu}(0)| p(\vec{0},\sigma_p)\rangle.
\end{eqnarray}
The matrix element in the r.h.s. of (\ref{label6.14}) is related to
the matrix element of the axial--vector current defining the
$\beta$--decay of the neutron \cite{MN79,IZ80a}
\begin{eqnarray}\label{label6.15}
\langle n(-\vec{q},\sigma)|J^{1-i2}_{5\mu}(0)|
p(\vec{0},\sigma_p)\rangle = g_A\,\bar{u}_n(-\vec{q},\sigma)
\gamma_{\mu} \gamma^5u(\vec{0},\sigma_p),
\end{eqnarray}
where $\bar{u}_n(-\vec{q},\sigma)$ and $u(\vec{0},\sigma_p)$ are
Dirac bispinors of the neutron and the proton.

Thus, the matrix element of the reaction $\pi^- + p \to n + \gamma$ is
determined by
\begin{eqnarray}\label{label6.16}
M(\pi^-(\vec{k}\,) p(-\vec{k}\,) \to n \gamma) =
-\,\sqrt{2\pi}\,\frac{ie g_A}{F_{\pi}}\,\bar{u}(-\vec{q},\sigma)
\gamma_{\mu} \gamma^5u(\vec{0},\sigma_p)\,e^{\mu}(\vec{q},\lambda).
\end{eqnarray}
The partial width of the decay $A_{\pi p} \to n + \gamma$ is equal to
\begin{eqnarray}\label{label6.17}
\Gamma(A_{\pi p} \to n \gamma) =
\alpha\,\frac{3}{4}\,\frac{g^2_A}{F^2_{\pi}}\,\frac{m_n}{~m_{\pi^-}}\,
\Big(1 - \frac{m^2_n}{(m_{\pi^-} + m_p)^2}\Big)|\Psi_{1s}(0)|^2 =
0.369\,{\rm eV}.
\end{eqnarray}
This value should be compared with the partial width of the decay
$A_{\pi p} \to n \pi^0$, which reads
\begin{eqnarray}\label{label6.18}
\Gamma(A_{\pi p} \to n \pi^0) =
\frac{8\pi}{9}\,\frac{p^*}{\mu}\,(a^{1/2}_0 -
a^{3/2}_0)^2\,|\Psi_{1s}(0)|^2 = 0.542\,{\rm eV}.
\end{eqnarray}
The Panofsky ratio $1/P$ is equal to
\begin{eqnarray}\label{label6.19}
\hspace{-0.3in}\frac{1}{P} =
\frac{27}{32}\frac{\alpha}{\pi}\frac{g^2_A}{F^2_{\pi}}\,
\frac{m_n}{~m_{\pi^-}}\frac{\mu}{p^*}\frac{1}{(a^{1/2}_0 -
a^{3/2}_0)^2}\Big(1 - \frac{m^2_n}{(m_{\pi^-} + m_p)^2}\Big) =
0.681\pm 0.048.
\end{eqnarray}
The theoretical value agrees with the experimental data $1/P = 0.647
\pm 0.004$ \cite{JS77}. The theoretical error is related to the errors of the
experimental values of the S--wave scattering lengths $a^{1/2}_0 -
a^{3/2}_0 = (0.2715 \pm 0.0095)\,m^{-1}_{\pi^-}$ \cite{PSI2}.

The cross section for the reaction $\pi^- + p \to n + \gamma$ at low
relative velocities of the $\pi^-p$ system $v$ is equal to
\begin{eqnarray}\label{label6.20}
\sigma(\pi^-p \to n\gamma) = \frac{432}{v}\,\,{\rm \mu barn}.
\end{eqnarray}
The result (\ref{label6.20}) agrees well with the theoretical estimate
given by Anderson and Fermi \cite{EF52}.

Now we are able to apply the technique developed above to the
calculation of the partial widths of the electromagnetic decay
channels of kaonic hydrogen $A_{Kp} \to \Lambda^0 + \gamma$ and
$A_{Kp} \to \Sigma^0 + \gamma$.

\subsection{Radiative decays of kaonic hydrogen}

Amplitudes of the decays $A_{Kp} \to \Lambda^0 + \gamma$ and $A_{Kp}
\to \Sigma^0 + \gamma$ we define in analogy to (\ref{label6.6}). This
gives
\begin{eqnarray}\label{label6.21}
M(A_{K p} \to \Lambda^0 \gamma) &=& \sqrt{\frac{1}{2\mu}}\int
\frac{d^3k}{(2\pi)^3}\,\sqrt{\frac{m_{K^-}m_p}{E_{K^-}(\vec{k}\,)
E_p(\vec{k}\,)}}\,\Phi_{1s}(\vec{k}\,)\,M(K^-(\vec{k}\,)
p(-\vec{k}\,) \to \Lambda^0 \gamma),\nonumber\\
M(A_{K p} \to \Sigma^0 \gamma) &=& \sqrt{\frac{1}{2\mu}}\int
\frac{d^3k}{(2\pi)^3}\,\sqrt{\frac{m_{K^-}m_p}{E_{K^-}(\vec{k}\,)
E_p(\vec{k}\,)}}\,\Phi_{1s}(\vec{k}\,)\,M(K^-(\vec{k}\,)
p(-\vec{k}\,) \to \Sigma^0 \gamma),\nonumber\\
&&
\end{eqnarray}
where $\mu = m_{K^-}m_p/(m_{K^-} + m_p) = 323.478\,{\rm MeV}$ is
the reduced mass of the $K^-p$ system and $\Phi_{1s}(\vec{k}\,)$ is
the wave function of the ground state of kaonic hydrogen in the
momentum representation.  

The amplitudes of the reactions $K^- + p \to \Lambda^0 + \gamma$ and
$K^- + p \to \Sigma^0 + \gamma$ read
\begin{eqnarray}\label{label6.22}
M(K^-(\vec{k}\,) p(-\vec{k}\,) \to \Lambda^0\gamma) &=&
\sqrt{4\pi}\,e\,\langle \Lambda^0(-\vec{q},\sigma)|J^{\rm e\ell
m}_{\mu}(0)|K^-(\vec{k}\,) p(- \vec{k},\sigma_p)\rangle\,
e^{\mu}(\vec{q},\lambda),\nonumber\\ M(K^-(\vec{k}\,) p(-\vec{k}\,)
\to \Sigma^0\gamma) &=& \sqrt{4\pi}\,e\,\langle
\Sigma^0(-\vec{q},\sigma)|J^{\rm e\ell m}_{\mu}(0)|K^-(\vec{k}\,) p(-
\vec{k},\sigma_p) \rangle\, e^{\mu}(\vec{q},\lambda).\nonumber\\ &&
\end{eqnarray}
The application of the reduction technique reduces the matrix elements
(\ref{label6.22}) to the form
\begin{eqnarray}\label{label6.23}
\hspace{-0.3in}&&\langle \Lambda^0(-\vec{q},\sigma)|J^{\rm e\ell
m}_{\mu}(0)|K^-(\vec{k}\,) p(-\vec{k},\sigma_p)\rangle =
\lim_{k^2_{K^-}\to m^2_{K^-}}i\int d^4x\,e^{\textstyle\,-ik_{K^-}\cdot
x}\,(\Box_x + m^2_{K^-})\nonumber\\ \hspace{-0.3in}&&\times\,\langle
\Lambda^0(-\vec{q},\sigma)| {\rm T}(J^{\rm e\ell m}_{\mu}(0)
K^{-\dagger}(x))| p(-\vec{k},\sigma_p)\rangle,\nonumber\\
\hspace{-0.3in}&&\langle \Sigma^0(-\vec{q},\sigma)|J^{\rm e\ell
m}_{\mu}(0)|K^-(\vec{k}\,) p(-\vec{k},\sigma_p)\rangle =
\lim_{k^2_{K^-}\to m^2_{K^-}}i\int d^4x\,e^{\textstyle\,-ik_{K^-}\cdot
x}\,(\Box_x + m^2_{K^-})\nonumber\\ \hspace{-0.3in}&&\times\,\langle
\Sigma^0(-\vec{q},\sigma)| {\rm T}(J^{\rm e\ell m}_{\mu}(0)
K^{-\dagger}(x))| p(-\vec{k},\sigma_p)\rangle.
\end{eqnarray}
The PCAC hypothesis allows to define the interpolating field
$K^{-\dagger}(x)$ in terms of the divergence of the axial--vector
current \cite{SA68,HP73}
\begin{eqnarray}\label{label6.24}
K^{-\dagger}(x) = \frac{1}{\sqrt{2}}\,\frac{1}{m^2_{K^-}F_K}\,
\partial^{\nu}J^{4-i5}_{5\nu}(x).
\end{eqnarray}
In the soft--kaon limit $k_{K^-} \to 0$ we obtain
\begin{eqnarray}\label{label6.25}
\hspace{-0.3in}&&\langle \Lambda^0(-\vec{q},\sigma)|J^{\rm e\ell
m}_{\mu}(0)|K^-(\vec{k}\,) p(-\vec{k},\sigma_p)\rangle
=\frac{i}{\sqrt{2}F_K}\langle \Lambda^0(-\vec{q},\sigma)|[J^{\rm e\ell
m}_{\mu}(0), Q^{4-i5}_5(0)]| p(\vec{0},\sigma_p)\rangle,\nonumber\\
\hspace{-0.3in}&&\langle \Sigma^0(-\vec{q},\sigma)|J^{\rm e\ell
m}_{\mu}(0)|K^-(\vec{k}\,) p(-\vec{k},\sigma_p)\rangle =
\frac{i}{\sqrt{2}F_K}\langle \Sigma^0(-\vec{q},\sigma)|[J^{\rm e\ell
m}_{\mu}(0), Q^{4-i5}_5(0)]| p(\vec{0},\sigma_p)\rangle.\nonumber\\
\hspace{-0.3in}&&
\end{eqnarray}
Using Gell--Mann's current algebra \cite{SA68,HP73} we transcribe the
r.h.s. of the matrix elements (\ref{label6.25}) into the form
\begin{eqnarray}\label{label6.26}
\hspace{-0.5in}&&\langle \Lambda^0(-\vec{q},\sigma)|J^{\rm e\ell
m}_{\mu}(0)|K^-(\vec{k}\,) p(-\vec{k},\sigma_p)\rangle
=-\,\frac{i}{\sqrt{2}F_K}\langle
\Lambda^0(-\vec{q},\sigma)|J^{4-i5}_{5\mu}(0)|
p(\vec{0},\sigma_p)\rangle,\nonumber\\
\hspace{-0.5in}&&\langle \Sigma^0(-\vec{q},\sigma)|J^{\rm e\ell
m}_{\mu}(0)|K^-(\vec{k}\,) p(-\vec{k},\sigma_p)\rangle = -\,
\frac{i}{\sqrt{2}F_K}\langle
\Sigma^0(-\vec{q},\sigma)|J^{4-i5}_{5\mu}(0)|
p(\vec{0},\sigma_p)\rangle,
\end{eqnarray}
where $F_K = 112.996\,{\rm MeV}$ is the PCAC constant of $K^{\pm}$
mesons \cite{DG00}.  The matrix elements of the axial--vector current
in the r.h.s. of (\ref{label6.26}) can be defined in analogy with
(\ref{label6.15})
\begin{eqnarray}\label{label6.27}
\langle \Lambda^0(-\vec{q},\sigma)|J^{4-i5}_{5\mu}(0)|
p(\vec{0},\sigma_p)\rangle &=& g^{\Lambda^0}_A
\bar{u}_{\Lambda^0}(-\vec{q},\sigma)\gamma_{\mu}\gamma^5
u(\vec{0},\sigma_p),\nonumber\\ \langle
\Sigma^0(-\vec{q},\sigma)|J^{4-i5}_{5\mu}(0)|
p(\vec{0},\sigma_p)\rangle &=& g^{\Sigma^0}_A
\bar{u}_{\Sigma^0}(-\vec{q},\sigma)\gamma_{\mu}\gamma^5
u(\vec{0},\sigma_p).
\end{eqnarray}
The partial widths of the decays $A_{Kp} \to \Lambda^0 \gamma$ and
$A_{Kp} \to \Sigma^0 \gamma$ are equal to
\begin{eqnarray}\label{label6.28}
\Gamma(A_{K p} \to \Lambda^0 \gamma) &=&
\alpha\,\frac{3}{4}\,\frac{(g^{\Lambda^0}_A)^2}{F^2_K}\,
\frac{m_{\Lambda^0}}{~m_{K^-}}\, \Big(1 -
\frac{m^2_{\Lambda^0}}{(m_{K^-} +
m_p)^2}\Big)|\Psi_{1s}(0)|^2,\nonumber\\ \Gamma(A_{K p} \to \Sigma^0
\gamma) &=& \alpha\,\frac{3}{4}\,\frac{(g^{\Sigma^0}_A)^2}{F^2_K}\,
\frac{m_{\Sigma^0}}{~m_{K^-}}\, \Big(1 -
\frac{m^2_{\Sigma^0}}{(m_{K^-} + m_p)^2}\Big)|\Psi_{1s}(0)|^2.
\end{eqnarray}
The coupling constant $g^{\Lambda^0}_A$ can be taken from the data on
the $\beta$--decay of the $\Lambda^0$--hyperon, $\Lambda^0 \to p + e^-
+ \bar{\nu}_e$: $g^{\Lambda^0}_A = 0.718 \pm 0.015$ \cite{DG00}. Due
to isospin invariance of strong interactions we can set
$g^{\Sigma^0}_A = g^{\Sigma^-}_A/\sqrt{2} = 0.240\pm 0.012$
\cite{JB68}, where $g^{\Sigma^-}_A = 0.340 \pm 0.017$ defines the
$\beta$--decay $\Sigma^- \to n + e^- + \bar{\nu}_e$ \cite{DG00}. As a
result we obtain the following numerical values of the partial widths
\begin{eqnarray}\label{label6.29}
\Gamma(A_{K p} \to \Lambda^0 \gamma) &=& (0.82 \pm 0.04)\,{\rm
eV},\nonumber\\ \Gamma(A_{K p} \to \Sigma^0 \gamma) &=& (0.08 \pm
0.01)\,{\rm eV},
\end{eqnarray}
where we have used $m_{\Lambda^0} = 1115.683\,{\rm MeV}$ and
$m_{\Sigma^0} = 1192.642\,{\rm MeV}$ \cite{DG00}. 

The parameter $X$, the inverse Panofsky ratio for kaonic hydrogen, is
equal to
\begin{eqnarray}\label{label6.30}
X &=& \frac{\Gamma(A_{Kp} \to \Lambda^0\gamma) + \Gamma(A_{Kp} \to
\Sigma^0\gamma)}{\Gamma_{1s}}
=\alpha\,\frac{3}{16\pi}\,\frac{1}{F^2_K}\,\frac{\mu}{~m_{K^-}
}\,\frac{1}{{\cal I}m\,f^{K^-p}_0(0)}\nonumber\\
&\times&\Big[(g^{\Lambda^0}_A)^2\,m_{\Lambda^0}\, \Big(1 -
\frac{m^2_{\Lambda^0}}{(m_{K^-} + m_p)^2}\Big) + (g^{\Sigma^0}_A)^2
m_{\Sigma^0}\, \Big(1 - \frac{m^2_{\Sigma^0}}{(m_{K^-} +
m_p)^2}\Big)\Big]=\nonumber\\ &=& (3.97 \pm 0.47)\times 10^{-3}.
\end{eqnarray}
Thus, the contribution of radiative decay channels $A_{Kp} \to
\Lambda^0 \gamma$ and $A_{Kp} \to \Sigma^0 \gamma$ to the width of the
ground state of kaonic hydrogen is less than 0.5$\%$.

The branching ratios ${\rm B}(A_{Kp} \to \Lambda^0\gamma) = (3.61\pm
0.43) \times 10^{-3}$ and ${\rm B}(A_{Kp} \to \Sigma^0\gamma) = (0.35
\pm 0.04)\times 10^{-3}$, obtained for the partial widths
(\ref{label6.29}) and the total width $\Gamma_{1s} = (227 \pm
27)\,{\rm eV}$ given by (\ref{label4.22}), are in qualitative
agreement with both theoretical values, predicted by Hamaie {\it et
al.}  \cite{TH96}, ${\rm B}(A_{Kp} \to \Lambda^0\gamma) = 4.72\times
10^{-3}$ and ${\rm B}(A_{Kp} \to \Sigma^0\gamma) = 2.43\times
10^{-3}$, and experimental values, ${\rm B}(A_{Kp} \to
\Lambda^0\gamma) = (0.86 \pm 0.12)\times 10^{-3}$ and ${\rm B}(A_{Kp}
\to \Sigma^0\gamma) = (1.44 \pm 0.23)\times 10^{-3}$ \cite{DW89}.

The branching ratio of the radiative decays of the $\Lambda(1405)$
resonance is equal to ${\rm B}(\Lambda(1405) \to \Lambda^0\gamma) +
{\rm B}(\Lambda(1405) \to \Sigma^0\gamma) = (0.13 \pm 0.03)\%$
\cite{DG00,HB91}. The data on radiative decays of the $\Sigma(1750)$
resonance are absent \cite{DG00}.

Hence, within an accuracy about 1$\%$ one can neglect the
contributions of radiative decay channels to the width of the ground
state of kaonic hydrogen.

\section{Conclusion}
\setcounter{equation}{0}

We have analysed the energy level displacement of the ground state of
kaonic hydrogen within a quantum field theoretic and relativistic
covariant approach. In our approach the energy level displacement of
the ground state of kaonic hydrogen is defined by the amplitude of the
reaction $K^- + p \to K^- + p$, weighted with the wave functions of
kaonic hydrogen in the ground state (\ref{label3.5}). It reads
$$
- \epsilon_{1s} + i\,\frac{\Gamma_{1s}}{2} =
  \frac{1}{4m_{K^-}m_p}\int\frac{d^3k}{(2\pi)^3}
  \int\frac{d^3q}{(2\pi)^3}\,
  \sqrt{\frac{m_{K^-}m_p}{E_{K^-}(\vec{k}\,)E_p(\vec{k}\,)}}\,
  \sqrt{\frac{m_{K^-}m_p}{E_{K^-}(\vec{q}\,) E_p(\vec{q}\,)}}
$$
$$
\times\Phi^{\dagger}_{1s}(\vec{k}\,)\,
  M(K^-(\vec{q}\,)p(-\vec{q},\sigma_p) \to
  K^-(\vec{k}\,)p(-\vec{k},\sigma_p))\,
  \Phi_{1s}(\vec{q}\,).\eqno(7.1)
$$
By virtue of the wave functions $\Phi^{\dagger}_{1s}(\vec{k}\,)$ and
$\Phi_{1s}(\vec{q}\,)$ the integrand is concentrated around momenta $k
\sim 1/a_B$ and $q \sim 1/a_B$, where $1/a_B = 2.361\,{\rm
MeV}$. Since typical momenta are much less than the masses of coupled
particles, $m_{K^-} \gg 1/a_B$ and $m_p \gg 1/a_B$, the zero--momentum
limit $k = q = 0$ turns out to be a good approximation\,\footnote{An
expansion in powers of the relative momenta should lead to the
corrections of order of powers of $\alpha$, i.e. the term of order
$O(\sqrt{kq})$ gives a correction of order $O(\alpha)$ and so on,
caused by Coulombic photons. We are planning to analyse these
corrections in our forthcoming publications.}. This results in the
well--known DGBT formula
$$
- \epsilon_{1s} + i\,\frac{\Gamma_{1s}}{2} =
  2\alpha^3\mu^2\,f^{K^-p}_0(0),\eqno(7.2)
$$
where $f^{K^-p}_0(0)$ is the partial S--wave amplitude of the reaction
$K^- + p \to K^- + p$ at threshold.

For the description of the amplitude $f^{K^-p}_0(0)$ we have suggested
the dominance of a smooth elastic background of low--energy $K^-p$
scattering and three resonances $\Lambda(1405)$, the $SU(3)_{\rm
flavour}$ singlet, and the $\Lambda(1800)$ and $\Sigma(1750)$, the
components of the $SU(3)_{\rm flavour}$ octet. These resonances
saturate the part of the amplitude which we have denoted as
$f^{K^-p}_0(0)_R$ (\ref{label3.10}).

The imaginary part of the amplitude $f^{K^-p}_0(0)_R$ is related to
inelastic channels $K^-p \to \Sigma^-\pi^+$,$ K^-p \to \Sigma^+\pi^-$,
$K^-p \to \Sigma^0\pi^0$ and $K^-p \to \Lambda^0\pi^0$, which are
fully described by the resonances $\Lambda(1405)$, $\Lambda(1800)$ and
$\Sigma(1750)$. 

For the analysis of the consistency of our approach, applied to the
description of inelastic channels $K^-p \to \Sigma^-\pi^+$,$ K^-p \to
\Sigma^+\pi^-$, $K^-p \to \Sigma^0\pi^0$ and $K^-p \to
\Lambda^0\pi^0$, we have used the experimental data $\gamma = 2.360
\pm 0.040$, $R_n = 0.189 \pm 0.015$ and $R_c = 0.664 \pm 0.011$
(\ref{label4.7}) on the ratios of the cross sections for the reactions
$K^-p \to \Sigma^-\pi^+$,$ K^-p \to \Sigma^+\pi^-$, $K^-p \to
\Sigma^0\pi^0$ and $K^-p \to \Lambda^0\pi^0$. We have found that in
our approach these experimental constraints are fulfilled within an
accuracy better than 6$\%$.

Moreover, we have shown that in our approach between three parameters
$\gamma$, $R_n$ and $R_c$ only two parameters are
independent. Assuming that these are $\gamma$ and $R_n$ we have
expressed $R_c$ in terms of $\gamma$ and $R_n$.  Using the
experimental values for the parameters $\gamma$ and $R_n$ we have
obtained $R_c = 0.626 \pm 0.007$ that agrees with experimental value
$R_c = 0.664 \pm 0.011$ within an accuracy better than 6$\%$.  Most
likely that the obtained agreement of our approach with experimental
data on $\gamma$, $R_n$ and $R_c$ is a consequence of the $SU(3)_{\rm
flavour}$ singlet--octet nature of the resonances $\Lambda(1405)$,
$\Lambda(1800)$ and $\Sigma(1750)$.

One of the consequences of the experimental data (\ref{label4.7}) on
the cross sections for inelastic channels of low--energy $K^-p$
scattering and the $SU(3)_{\rm flavour}$ singlet--octet nature of the
resonances $\Lambda(1405)$, $\Lambda(1800)$ and $\Sigma(1750)$ is a
suppression of the contribution of the $\Lambda(1800)$
resonance. Indeed, due to the experimental constraints
(\ref{label4.7}) the ratio of the coupling constants of the
antisymmetric and symmetric $SU(3)_{\rm flavour}$ phenomenological
$B_2BP$ interactions, $\alpha_2 = f_2/g_2$, turns out to be very close
to $- 1/3$. Since the coupling constant of the $\Lambda(1800)$
resonance with the $\bar{K}N$ pairs is proportional to $(1 +
3\alpha_2)$, it decouples from the $\bar{K}N$ system for $\alpha_2 = -
1/3$.

For the numerical analysis of the amplitude of $K^-p$ scattering near
threshold we have used the recommended values for the masses and total
widths of the resonances $\Lambda(1405)$ and $\Sigma(1750)$:
$m_{\Lambda(1405)} = 1406\,{\rm MeV}$, $\Gamma_{\Lambda(1405)} =
50\,{\rm MeV}$ and $m_{\Sigma(1750)} = 1750\,{\rm MeV}$ and
$\Gamma_{\Sigma(1750)} = 90\,{\rm MeV}$. This has given the following
value of the resonant part of the amplitude of $K^-p$ scattering near
threshold
$$
f^{K^-p}_0(0)_R = (-\,0.154 \pm 0.009) + i\,(0.269 \pm 0.032)\;{\rm
fm}.\eqno(7.3)
$$
Since the smooth elastic background should be fully real, the
imaginary part of $f^{K^-p}_0(0)_R$ coincides with the imaginary part
of the S--wave amplitude $f^{K^-p}_0(0)$ of $K^-p$ scattering near
threshold. As a result it should fit the experimental data on the
width of the energy level of kaonic hydrogen in the ground
state. Using the DGBT formula, which is the non--relativistic
reduction of our formula (7.1), we have got the value $\Gamma^{\rm
th}_{1s} = (227 \pm 27)\;{\rm eV}$ fitting well the meanvalue of the
experimental data by the DEAR Collaboration $\Gamma_{1s} = (213 \pm
138)\;{\rm eV}$ \cite{DEAR4}.

The shift $\epsilon_{1s}$ of the energy level of kaonic hydrogen in
the ground state is defined by the S--wave scattering length
$a^{K^-p}_0$ of $K^-p$ scattering. In our approach $a^{K^-p}_0$, the
real part of the amplitude $f^{K^-p}_0(0)$, is determined by the sum
of the contributions of the resonances and a smooth elastic
background: $a^{K^-p}_0 = {\cal R}e\,f^{K^-p}_0(0) = {\cal
R}e\,f^{K^-p}_0(0)_R + A^{K^-p}_B$.

We have calculated the contribution of the smooth elastic background
within the Effective quark model with chiral $U(3)\times U(3)$
symmetry: $A^{K^-p}_B = (-\,0.328 \pm 0.033)\;{\rm fm}$. This gives
the S--wave scattering length $a^{K^-p} = (-0.482 \pm 0.034)\,{\rm
fm}$ and the shift of the energy level of the ground state of kaonic
hydrogen $\epsilon^{\rm th}_{1s} = (203 \pm 15)\;{\rm eV}$, which fits
well the experimental data $\epsilon^{\exp}_{1s} = (183 \pm 62)\;{\rm
eV}$ by the DEAR Collaboration \cite{DEAR4}.

At the hadronic level we have calculated the parameter $A^{K^-p}_B$ in
terms of the contribution coming from all hadron exchanges taken at
leading order in ChPT, described by Effective Chiral Lagrangians, and
scalar mesons $a_0(980)$ and $f_0(980)$ having an exotic
$qq\bar{q}\bar{q}$ (or $\bar{K}K$ molecule) structure. Due to the lack
of information about $a_0(980)NN$ and $f_0(980)NN$ coupling constants,
the parameter $A^{K^-p}_B$ has been found dependent on an arbitrary
parameter $\xi$. Comparing this expression with that obtained at the
quark level we have estimated $\xi = 1.2 \pm 0.1$. Of course, an
additional information about the value of $\xi$ can be extracted from
the analysis of the contributions of the $a_0(980)$ and $f_0(980)$
mesons to the reactions of $\bar{K}N$ interaction at transferred
momenta of order of $1\,{\rm GeV}$.

Thus, in our approach the S--wave amplitude $f^{K^-p}_0(0)$ of $K^-p$
scattering near threshold is equal to
$$
f^{K^-p}_0(0) = (- 0.482 \pm 0.034) + i\,(0.269 \pm 0.032)\;{\rm
fm}.\eqno(7.4)
$$
This leads to the following theoretical prediction for the energy
level displacement of the ground state of kaonic hydrogen
$$
- \epsilon^{\rm th}_{1s} + i\,\frac{\Gamma^{\rm th}_{1s}}{2} = (- 203
  \pm 15) + i\,(113 \pm 14)\,{\rm eV},\eqno(7.5)
$$
which fits well the experimental data by the DEAR Collaboration
\cite{DEAR4}
$$
- \epsilon^{\exp}_{1s} + i\,\frac{\Gamma^{\exp}_{1s}}{2} = (- 183 \pm
  62) + i\,(106\pm 69)\,{\rm eV}.\eqno(7.6)
$$
The calculation of the partial widths of the radiative decay channels
of pionic and kaonic hydrogen we have carried out within the
soft--pion and soft--kaon technique \cite{SA68,HP73}\,\footnote{A
constituent quark--diagram technique for the derivation of the
soft--pion and soft--kaon low--energy theorems has been elaborated by
Natalia Troitskaya in \cite{NT90} (see also \cite{AI92,AI91}).}. We
have shown that for pionic hydrogen the partial width of the decay
$A_{\pi p} \to n + \gamma$ gives the Panofsky ratio
$$
\frac{1}{P} = \frac{\Gamma(A_{\pi p} \to n \gamma)}{\Gamma(A_{\pi p}
\to n\pi^0)} = 0.681\pm 0.048 \eqno(7.7)
$$
agreeing well with the experimental value $1/P = 0.647 \pm 0.004$
\cite{JS77}.

Unlike pionic hydrogen, where the radiative decay $A_{\pi p} \to n +
\gamma$ gives a contribution of about 65$\%$, the contribution of the
radiative decay channels $A_{Kp} \to \Lambda^0 + \gamma$ and $A_{Kp}
\to \Sigma^0 + \gamma$ is less than 1$\%$.  The theoretical
predictions for the sum of the branching ratios of the radiative decay
channels of the $\Lambda(1405)$ resonance makes up $(0.13 \pm 0.03)\%$
\cite{DG00,HB91}\,\footnote{Theoretical and experimental data on the
radiative decays of the $\Sigma(1750)$ resonance are absent
\cite{DG00}.}. 

Thus, the value of the parameter $X$, supplemented by the contribution
of the radiative decays of the $\Lambda(1405)$ resonance, does not
exceed 1$\%$. Since both theoretical and experimental accuracy of the
definition of the energy level displacement of the ground state of
kaonic hydrogen are worse than $1\%$, one can neglect the contribution
of the electromagnetic decay channels of kaonic hydrogen to the total
width $\Gamma_{1s}$.

Thus, we can argue that strong low--energy $\bar{K}N$ interactions
define fully the experimental value of the energy level displacement
of kaonic hydrogen measured by the DEAR Collaboration\,\footnote{A
tangible contribution about 50$\%$ to the paramter $X$, coming from
the isospin--breaking and electromagnetic interactions to the
amplitude of low--energy $K^-p$ scattering through the intermediate
$\bar{K}^0n$ state $K^-p \to \bar{K}^0n \to K^-p$, has been recently
pointed out by Rusetsky \cite{AR03}.}.

An agreement of our theoretical predictions for the energy level
displacement of the ground state of kaonic hydrogen (7.5) with the
experimental data by Iwasaki {\it et al.}  (the KEK experiment)
\cite{IW97}
$$
- \epsilon^{\exp}_{1s} + i\,\frac{\Gamma^{\exp}_{1s}}{2} = (- 323 \pm
  63 \pm 11) + i\,(204 \pm 104 \pm 50)\,{\rm eV}.\eqno(7.8)
$$
seems to be only qualitative.

We would like to emphasize that the new data on the energy level
displacement have been obtained by the DEAR Collaboration due to a
significant improvement of the experimental technique and methodics of
the extraction of the energy level displacement of kaonic hydrogen
from the data on the $np\to 1s$ transitions, where $np$ is an excited
state of kaonic hydrogen \cite{DEAR4}.

\section*{Acknowledgement}

We are grateful to Georgii Shestakov for clarification of exotic
properties of the $a_0(980)$ and $f_0(980)$ mesons as the four--quark
(or $\bar{K}K$ molecule) states and John Donoghue for discussions. We
appreciate discussions with J\"urg Gasser and Akaki Rusetsky.

We thank Torleif Ericson for the discussions of our approach to the
description of the Panofsky ratio for pionic hydrogen and Evgeni
Kolomeitsev for the discussions of low--energy theorems for $K^-p$
scattering.

The results obtained in this paper have been reported at the Workshop
on {\it CHIRAL DYNAMICS} at University of Bonn, 8--13 September 2003,
Germany and at the Workshop on {\it HADATOM03} at ECT$^*$ in Trento,
12--18 October 2003, Italy  \cite{CD03,AI03}.

\newpage

\section*{Appendix. Calculation of $A^{K^-p}_B$ within Effective 
quark model with chiral $U(3)\times U(3)$ symmetry}

Using the expression for the external sources $\eta_p(x_2)$ and
$\bar{\eta}_p(x_3)$, given by (\ref{label5.19}), and substituting them
in (\ref{label5.22}) we obtain
$$
M(K^- p \to K^- p) = i\,\frac{1}{4}\,g^2_{\rm
B}\,g^2_K\,\varepsilon^{i\,'j\,'k\,'}\,\varepsilon^{ijk}\int d^4x_1
d^4x_2 d^4x_3\,e^{\textstyle i\,q\,' \cdot x_1 + ip\,'\cdot x_2 -
ip\cdot x_3}
$$
$$
\times\,\bar{u}(p\,',\sigma\,'\,)_a (i\gamma^5)_{a_1b_1}
(C\gamma^{\mu})_{a_2b_2} (\gamma_{\mu}\gamma^5)_{a c_2}
(\gamma_{\nu}\gamma^5)_{c_3 b}(\gamma_{\nu}C)_{a_3b_3}
(i\gamma^5)_{a_4b_4}u(p,\sigma)_b
$$
$$
\hspace{-0.1in}\times\langle 0|{\rm
T}(\bar{u}_{\ell}(x_1)_{a_1}s_{\ell}(x_1)_{b_1}
u_{i\,'}(x_2)_{a_2}u_{j\,'}(x_2)_{b_2}
d_{k\,'}(x_2)_{c_2}\bar{d}_i(x_3)_{c_3}\bar{u}_j(x_3)_{a_3}
\bar{u}_k(x_3)_{b_3} \bar{s}_t(0)_{a_4} u_t(0)_{b_4})|0\rangle_c,
\eqno({\rm A}.1)
$$
where the index $c$ stands for the abbreviation {\it connected}.

Making contractions of the $d$-- and $s$--quark field operators we
reduce the r.h.s of ({\rm A}.1) to the form
$$
M(K^- p \to K^- p) = i\,\frac{1}{4}\,g^2_{\rm
B}\,g^2_K\,\varepsilon^{i i\,'j\,'}\,\varepsilon^{ijk}\int d^4x_1
d^4x_2 d^4x_3\,e^{\textstyle i\,q\,' \cdot x_1 + ip\,'\cdot x_2 -
ip\cdot x_3}
$$
$$
\times\,\bar{u}(p\,',\sigma\,'\,)_a (i\gamma^5)_{a_1b_1}
(C\gamma^{\mu})_{a_2b_2} (\gamma_{\mu}\gamma^5)_{a c_2}
(\gamma_{\nu}\gamma^5)_{c_3 b}(\gamma_{\nu}C)_{a_3b_3}
(i\gamma^5)_{a_4b_4}u(p,\sigma)_b
$$
$$
\times\,(-i)\,S^{(s)}_F(x_1)_{b_1 a_4}\,(-i)\,S^{(d)}_F(x_2 -
x_3)_{c_2 c_3}
$$
$$
\times\,\langle 0|{\rm T}(\bar{u}_{\ell}(x_1)_{a_1}
u_{i\,'}(x_2)_{a_2}u_{j\,'}(x_2)_{b_2} \bar{u}_j(x_3)_{a_3}
\bar{u}_k(x_3)_{b_3} u_{\ell}(0)_{b_4})|0\rangle_c, \eqno({\rm A}.2)
$$
The requirement to deal with only {\it connected} quark diagrams
prohibits the contraction of the $u$--quark field operators
$\bar{u}_{\ell}(x_1)_{a_1}$ and $u_{\ell}(0)_{b_4}$. The result reads
$$
M(K^- p \to K^- p) = 3\,g^2_{\rm B}\,g^2_K\int d^4x_1 d^4x_2
d^4x_3\,e^{\textstyle i\,q\,' \cdot x_1 + ip\,'\cdot x_2 - ip\cdot
x_3}
$$
$$
\times\,\bar{u}(p\,',\sigma\,'\,)_a (\gamma^5)_{a_1b_1}
(C\gamma^{\mu})_{a_2b_2} (\gamma_{\mu}\gamma^5)_{a c_2}
(\gamma_{\nu}\gamma^5)_{c_3 b}(\gamma_{\nu}C)_{a_3b_3}
(\gamma^5)_{a_4b_4}u(p,\sigma)_b
$$
$$
\times\,\S^{(s)}_F(x_1)_{b_1 a_4}\,S^{(d)}_F(x_2 - x_3)_{c_2 c_3}
S^{(u)}_F(x_2 - x_1)_{a_2a_1} S^{(u)}_F(x_2 - x_3)_{b_2a_3}
S^{(u)}_F(-x_3)_{b_4b_3}.\eqno({\rm A}.3)
$$
Summing over the indices we end up with the expression
$$
M(K^- p \to K^- p) = 3\,g^2_{\rm B}\,g^2_K\int d^4x_1 d^4x_2
d^4x_3\,e^{\textstyle i\,q\,' \cdot x_1 + ip\,'\cdot x_2 - ip\cdot
x_3}
$$
$$
\times\,\bar{u}(p\,',\sigma\,'\,)\gamma^{\mu}\gamma^5S^{(d)}_F(x_2 -
x_3)\gamma^{\nu}\gamma^5 u(p,\sigma)
$$
$$
\times\,{\rm tr}\{\gamma^5 S^{(s)}_F(x_1) \gamma^5 S^{(u)}_F(- x_3)
C^T\gamma^T_{\nu} S^{(u)}_F(x_2 - x_3)^T \gamma^T_{\mu}C^T
S^{(u)}_F(x_2 - x_1)\}.\eqno({\rm A}.4)
$$
Using the relations
$$
C^T\gamma^T_{\nu} S^{(u)}_F(x_2 - x_3)^T \gamma^T_{\mu}C^T = -
\gamma_{\nu} S^{(u)}_F(x_3 - x_2) \gamma_{\mu} \eqno({\rm A}.5)
$$
we transcribe the r.h.s. of ({\rm A}.4) into the form
$$
M(K^- p \to K^- p) = -\,3\,g^2_{\rm B}\,g^2_K\int d^4x_1
d^4x_2 d^4x_3\,e^{\textstyle i\,q\,' \cdot x_1 + ip\,'\cdot x_2 -
ip\cdot x_3}
$$
$$
\times\,\bar{u}(p\,',\sigma\,'\,)\gamma^{\mu}\gamma^5S^{(d)}_F(x_2 -
x_3)\gamma^{\nu}\gamma^5 u(p,\sigma)
$$
$$
\times\,\Big[{\rm tr}\{\gamma^5 S^{(s)}_F(x_1) \gamma^5 S^{(u)}_F(-
x_3) \gamma_{\nu} S^{(u)}_F(x_3 - x_2)\gamma_{\mu} S^{(u)}_F(x_2 -
x_1)\}. \eqno({\rm A}.6)
$$
In the momentum representation the r.h.s. of ({\rm A}.6) reads
$$
M(K^- p \to K^- p) = 3\,g^2_{\rm B}\,g^2_K\Big[\int
\frac{d^4k_1}{(2\pi)^4 i}\frac{d^4k_2}{(2\pi)^4 i}\,
\bar{u}(p\,',\sigma\,'\,)\gamma^{\mu}\gamma^5\frac{1}{m_d -
\hat{k}_1}\gamma^{\nu}\gamma^5 u(p,\sigma)
$$
$$
\times\,{\rm tr}\Big\{\gamma^5 \frac{1}{m_s - \hat{k}_2}\gamma^5
\frac{1}{m_u - \hat{k}_2 + \hat{q}}\gamma_{\nu}\frac{1}{m_u -
\hat{k}_2 - \hat{k}_1 + \hat{p} + \hat{q}}\gamma_{\mu}\frac{1}{m_u -
\hat{k}_2 + \hat{q}\,'}\Big\}.\eqno({\rm A}.7)
$$
The result of the calculation of momentum integrals within the
procedure accepted in the Effective quark model with chiral
$U(3)\times U(3)$ symmetry \cite{AI99}--\cite{AI92} is equal to
$$
M(K^- p \to K^- p) = \frac{g^2_B}{8\pi^2}\,\frac{\langle
\bar{q}q\rangle}{F^2_K}\,\mu\,\frac{m_s + m}{m_s -
m}\,\Big[m^2_s\,{\ell n}\Big(1 + \frac{\Lambda^2_{\chi}}{m^2_s}\Big) -
m^2\,{\ell n}\Big(1 +
\frac{\Lambda^2_{\chi}}{m^2}\Big)\Big],\eqno({\rm A}.8)
$$
where $\langle \bar{q}q\rangle = - (252.630 {\rm MeV})^3$ is the quark
condensate, $\Lambda_{\chi} = 940\,{\rm MeV}$ is the scale of the
spontaneous breaking of chiral symmetry \cite{AI99,AI92}. The parameter
$A^{K^-p}_B$ is given by
$$
A^{K^-p}_B = \frac{M(K^- p \to K^- p)}{8\pi (m_{K^-} + m_p)} =
$$
$$
= \frac{g^2_B}{64\pi^3}\,\frac{\langle
\bar{q}q\rangle}{F^2_K}\,\frac{\mu}{m_{K^-} + m_p}\,\frac{m_s + m}{m_s
- m}\,\Big[m^2_s\,{\ell n}\Big(1 + \frac{\Lambda^2_{\chi}}{m^2_s}\Big)
- m^2\,{\ell n}\Big(1 + \frac{\Lambda^2_{\chi}}{m^2}\Big)\Big] = -
0.328\,{\rm fm}. \eqno({\rm A}.9)
$$
A theoretical accuracy of this result is about of 10$\%$
\cite{AI99}--\cite{AI92} and \cite{AI80}.

\end{document}